\documentclass[12pt,a4paper]{article}
\pdfoutput=1
\usepackage{jheppub}

\usepackage{bm}
\usepackage{ifsym}
\usepackage{amsthm}
\usepackage{amsfonts}
\usepackage{ccaption}
\usepackage{mathrsfs}
\usepackage{booktabs}

\makeatletter
\@addtoreset{equation}{section}

\makeatletter
\renewcommand\section{\@startsection {section}{1}{\z@}%
                                   {-3.5ex \@plus -1ex \@minus -.2ex}
                                   {2.3ex \@plus.2ex}%
                                   {\normalfont\large\bfseries}}
\renewcommand\subsection{\@startsection{subsection}{2}{\z@}%
                                     {-3.25ex\@plus -1ex \@minus -.2ex}%
                                     {1.5ex \@plus .2ex}%
                                     {\normalfont\bfseries}}

  \captionnamefont{\bfseries}
  \captiontitlefont{\small\sffamily}
  \captiondelim{: }
  \hangcaption

\def\sec#1{\S\ref{#1}}
\def\fig#1{Fig.\,\ref{#1}}
\def\req#1{(\ref{#1})}

\def\thuss{\qquad \Longrightarrow \qquad}

\definecolor{rust}{rgb}{0.8,0.2,0.2}
\definecolor{green}{rgb}{0.1,0.8,0.2}

\def\AdS#1{AdS$_{#1}$}
\def\SAdS#1{Schwarzschild-AdS$_{#1}$}

\def\bulk{{\cal M}}

\def\ra{{\cal A}}
\def\rac{{\cal A}^c}
\def\raal{{\cal A}_\alpha}
\def\SA{S_\ra}
\def\SAc{S_{\rac}}
\def\Srho{S_{\rho_\Sigma}}
\def\dS{\delta S_\ra}
\def\entsurf{\partial {\cal A}}

\def\extr#1{{\mathfrak E}_{#1}}
\def\mins#1{{\mathfrak M}_{#1}}

\def\trans{{\cal X}}

\def\cwedge{\blacklozenge_{{\cal A}}}

\def\domd{\Diamond_{\cal A}}

\def\ph{\theta}
\def\phA{\theta_\infty}

\def\rh{r_+}

\def\CO{{\cal O}}
\def\alphaCW{\alpha_{\blacklozenge}}

\newcommand{\X}{{\mathfrak n}}
\newcommand{\M}{\mathfrak{m}}
\newcommand{\sff}{K}
\newcommand{\dee}{\mathrm{d}}
\newcommand{\traceK}{{\tt k}}
\newcommand{\ft}{{\tt t}}

\title{Holographic entanglement plateaux}

\author{Veronika E. Hubeny$^a$,}
\author{ Henry Maxfield$^a$,}
\author{  Mukund Rangamani$^a$}
\author{\& Erik Tonni$^b$}

\affiliation[a]{ Centre for Particle Theory \& Department of Mathematical Sciences,\\
Science Laboratories, South Road, Durham DH1 3LE, UK.}
\affiliation[b]{SISSA and INFN, via Bonomea 265,  34136, Trieste, Italy}
\emailAdd{veronika.hubeny@durham.ac.uk}
\emailAdd{h.d.maxfield@durham.ac.uk}
\emailAdd{mukund.rangamani@durham.ac.uk}
\emailAdd{erik.tonni@sissa.it}

\abstract{
We consider the entanglement entropy for holographic field theories in finite volume. We show  that the Araki-Lieb inequality is saturated for large enough subregions, implying that the thermal entropy can be recovered from the knowledge of the region and its complement.
 We observe 
 that this actually is forced upon us in holographic settings due to non-trivial features of the causal wedges associated with a given boundary region. 
 In the process, we present an infinite set of extremal surfaces in Schwarzschild-AdS geometry anchored on a given entangling surface.
 We also offer some speculations regarding the homology constraint required for computing holographic entanglement entropy. 
 } 


\begin{document}
\begin{flushright} \small{DCPT-13/23} \end{flushright}

\maketitle

\flushbottom

\section{Introduction}
\label{s:intro}

Entanglement is one of the most non-classical features of quantum mechanics and much effort has been expended in trying to figure out a quantitative measure for the amount of entanglement inherent in a quantum state. In local quantum field theories one such measure is provided by the entanglement entropy $\SA$ associated with a specified (spacelike) region $\ra$ located on a Cauchy slice $\Sigma$. This quantity is defined to be the von Neumann entropy of the reduced density matrix $\rho_\ra$ obtained by integrating out the degrees of freedom in the complementary region $\rac$. Famously the entanglement entropy is UV divergent, with the leading divergence being given by the area of the entangling surface $\entsurf$ (the boundary  of $\ra$). However, the finite part of the entanglement entropy contains non-trivial information about the quantum state and in certain cases serves as a novel order parameter.
As usual, it is this finite part which we shall consider.

Our interest is in understanding the behaviour of entanglement entropy for field theories defined on compact spatial geometries; so we take $\Sigma$ to be a compact Riemannian  manifold (typically ${\bf S}^{d-1}$). One of the questions which we wish to address is whether the entanglement entropy for a fixed state of the field on $\Sigma$ is a smooth function of the (geometrical attributes of the) region $\ra$. We envisage considering a family of {\em smooth} regions $\raal$ characterized by some parameter $\alpha$, which is a proxy for the relative size of $\ra \subset \Sigma$. 
The main question we want to ask is whether $S_\ra(\alpha)$ is smooth under changing $\alpha$.
 
For finite systems, analogy with statistical mechanics suggests that this would indeed be true. There should be no room for any discontinuity in $S_\ra(\alpha)$ since the reduced density matrix will change analytically with $\alpha$. The place where this is expected to break down is when we attain some analog of the thermodynamic limit, i.e., when the number of degrees of freedom involved gets  large. So a natural place to look for non-smooth behaviour is in the dynamics of large $N$ field theories on compact spacetimes in the planar limit.  This naturally motivates the study of field theories which can be captured holographically by gravitational dynamics using the gauge/gravity correspondence. We will refer to these holographic field theories as large $c$ (central charge) theories; for conventional non-abelian gauge theories $c\sim N^2$.

In past few years much progress has been made in understanding entanglement entropy in large $N$ field theories (at strong coupling) thanks to the seminal work of Ryu \& Takayanagi [RT] \cite{Ryu:2006bv,Ryu:2006ef}, who gave a very simple geometric prescription for computing $S_\ra$ for static states in terms of the area of a bulk minimal surface anchored on $\entsurf$. This prescription was extended by Hubeny, Rangamani, and Takayanagi [HRT] to arbitrary time-dependent states in  \cite{Hubeny:2007xt} where one considers extremal surfaces, which can also be related to light-sheets discussed in  covariant entropy bound context  \cite{Bousso:2002ju}.
There have been many attempts to derive the prescription from first principles; the first was made in \cite{Fursaev:2006ih} which was critically examined in \cite{Headrick:2010zt}. More recently, \cite{Casini:2011kv} gave a nice argument deriving the prescription for a special class of states (conformally invariant vacuum) and spherically symmetric regions, by converting the reduced density matrix $\rho_\ra$ to a thermal density matrix. A local version of this argument has been made recently in \cite{Lewkowycz:2013nqa} and together with the results of \cite{Hartman:2013mia, Faulkner:2013yia} goes quite a way in establishing the holographic prescription of \cite{Ryu:2006bv} for static states. Given these developments, it is apposite to take stock of the implications of the holographic entanglement entropy prescription for the question we have in mind.

Let us first record some basic facts about $\SA$ which we will use to investigate the behaviour of $\SA(\alpha)$. It is a well known fact that when the total state of the field theory is pure,  the entanglement entropy of a given region and its complement are the same: $\SA = \SAc$. This however ceases to be true when the entire system is itself in a density matrix $\rho_\Sigma$. To measure the deviation from purity of the system we could monitor the difference $\dS = \SA - \SAc$. From our perspective this quantity has some useful advantages. Firstly, it is finite since the divergent contributions which are given in terms of intrinsic and extrinsic geometry of the entangling surface $\entsurf$ cancel. Secondly, it is bounded from above by the von Neumann entropy of the entire density matrix $\rho_\Sigma$, by the Araki-Lieb inequality \cite{Araki:1970ba}:
\begin{align}
\mid \dS \mid\; \equiv\; \mid  \SA-\SAc \mid \;\leq\; S_{\ra \cup \rac} = S_{\rho_{\Sigma}} \,.
\label{}
\end{align}

So one way to phrase our original question is to ask whether $\dS(\alpha)$ is a smooth function of $\alpha \in [0,1]$ which we take to be a suitable function of the ratio of $\text{Vol}(\ra)/\text{Vol}(\Sigma)$
such that
$\dS$ is an odd-function around $\alpha = \frac{1}{2}$ and becomes the total entropy $\pm S_{\rho_{\Sigma}}$ at $\alpha = 0,1$. The issue we want to focus on is whether there is any discontinuity either in $\SA(\alpha)$ or its derivatives as we vary $\alpha$. We will argue here that for large $c$ field theories $\SA(\alpha)$  (given by the RT prescription) has to be continuous for time-independent (static)
density matrices $\rho_\Sigma$.  However, there can be non-trivial behaviour in $\partial_\alpha \SA(\alpha)$: we will exhibit explicit examples where the function $\SA(\alpha)$ is continuous but fails to be differentiable.\footnote{
In fact, as we indicate in \sec{s:discuss}, in time dependent examples the situation may be much more intricate  and will be discussed elsewhere \cite{Hubeny:2013uq}.}$^{,}$\footnote{
Earlier studies of holographic entanglement entropy have indeed revealed examples where $S(\alpha)$ undergoes a first order phase transition, i.e., $\partial_\alpha S(\alpha)$ is discontinuous due to change in the nature of the minimal surfaces \cite{Nishioka:2006gr, Klebanov:2007ws}. For example the recent analysis of \cite{Liu:2012eea} exhibits this phenomenon arising due to the minimal surface changing topology (even in causally trivial spacetimes) as we vary $\alpha$.}
It is important to distinguish this from situations where the total density matrix itself varies; one can certainly have entanglement entropy of a fixed-size region which is discontinuous e.g.\ as a function of temperature \cite{Albash:2012pd, Belin:2013dva},  
as is quite familiar from the simpler example of Hawking-Page transition.  Here we fix the total density matrix (which has the bulk equivalent of fixing the spacetime) and consider the entanglement entropy as function of the region.

To understand the potential issues, we need to explain one key feature of the holographic prescription of \cite{Ryu:2006bv, Hubeny:2007xt}. To compute $\SA$, we find extremal surfaces\footnote{
We use the notation $\extr{\ra}$ for generic extremal surfaces and indicate minimal surfaces relevant for static geometries by $\mins{\ra}$. While we review the various assertions in terms of extremal surfaces we will for the most part (until \sec{s:homol}) only consider minimal surfaces in this paper.}  $\extr{\ra}$ in the bulk spacetime ${\cal M}$ which are anchored on $\entsurf$; in the asymptotically AdS spacetimes we consider, we demand that $\partial \extr{\ra} = \entsurf$. However, there can be multiple such  surfaces in a given spacetime. We are instructed to restrict attention to extremal surfaces $\extr{\ra}$ which are homologous to the boundary region $\ra$ under consideration \cite{Headrick:2007km} and from the set of such surfaces pick the one with smallest area. To wit,
\begin{align}
\SA = \min_X \frac{\text{Area}(\extr{\ra})}{4\, G_N} \,, \qquad 	X =  \extr{}:
	\begin{cases}
		& \; \partial \extr{\ra}\equiv \extr{\ra}\big|_{\partial {\cal M}} = \entsurf
		\\
		&
		\exists \;{\cal R} \subset {\cal M} : \partial {\cal R}= \extr{\ra} \cup \ra \\
		\end{cases} 
\label{HRTprop}
\end{align}
where the region ${\cal R}$ is a bulk co-dimension one smooth  surface (in the $d+1$ dimensional spacetime ${\cal M}$) which is bounded by the extremal surface $\extr{\ra}$ and the region $\ra$ on the boundary.

It was appreciated already in \cite{Headrick:2007km} that the homology constraint is crucial for the Araki-Lieb inequality to be satisfied. The issue was further elaborated in \cite{Azeyanagi:2007bj} where 1+1 dimensional CFTs on a torus were considered (see also \cite{Blanco:2013joa} for a recent discussion). In particular, its effects are most acutely felt when $\rho_\Sigma$ is a density matrix, for then we anticipate the bulk spacetime to have a horizon \cite{Freivogel:2005qh}.\footnote{
This statement presumes that  the von Neumann entropy of $\rho_\Sigma$ scales like the central charge of the field theory.} In static spacetimes this implies a non-trivial topology in the bulk when restricted to a constant time slice, which can be easily intuited by considering the Euclidean section of the geometry.

The homology constraint, being non-local from the boundary perspective, allows non-trivial behaviour in the nature of the extremal surfaces that are admissible for the problem. Indeed as we will  show explicitly, there are many simple examples where one has multiple extremal surfaces and only some of them are homologous to the boundary region in question.  In fact, the most bizarre aspect of our analysis is that for certain choices of boundary regions there are no connected minimal surfaces satisfying the homology constraint: one is forced into considering disconnected surfaces.\footnote{
The exchange of dominance between connected and disconnected surfaces in confining backgrounds for field theories on non-compact geometries ${\mathbb R}^{1,d-2} \times {\bf S}^1$ has been well studied in the context of holographic entanglement entropy, cf., \cite{Nishioka:2006gr, Klebanov:2007ws} for initial work. We will be however be considering field theories on compact spatial volumes.}
 Multiply connected extremal surfaces in turn imply that one can have novel behaviour in $\SA$ or equivalently in $\dS$; we go on to show that in static spacetimes these indicate that $\dS(\alpha)$ is continuous but not differentiable, exhibiting explicit examples involving global AdS black hole geometries. We argue that the lack of differentiability is the worst it gets for $\dS$ in static spacetimes, proving that $\dS$ has to be a continuous function of $\alpha$. Furthermore, when we are forced onto the branch of disconnected extremal surfaces, it is easy to establish that $\dS = \Srho$, i.e., the Araki-Lieb inequality is saturated; we refer to this phenomenon as {\em entanglement plateau}.\footnote{ This phenomenon like many others encountered in holographic duals is a feature of large $c$ theories. At finite central charge we cannot have any sharp plateaux; we thank Hong Liu for discussions on this issue.}

 While  in simple examples one can establish entanglement plateaux  by explicit construction, it is interesting to examine when it should happen on general grounds. Curiously, it is easy to provide a bound, though in a somewhat roundabout manner using machinery outside the extremal surface technology. The  necessary concept is geometric and has to do with the behaviour of the causal wedge associated with $\ra$. These objects were studied in the context of `causal holographic information' in \cite{Hubeny:2012wa} from a very different perspective (the motivation being to characterize the minimal amount of holographic information in the reduced density matrix). Using the topology of the causal wedge one can establish criteria for when the extremal surfaces $\extr{\ra}$ become disconnected. The precise statement and its proof will appear elsewhere \cite{Hubeny:2013fk}, but we will flesh out the physical aspects of the argument in what follows. Suffice to say for now that we find the interplay between causality and extremal surfaces extremely intriguing and believe that it points to some yet to be fathomed facet of holography. 
 
 The outline of the paper is as follows: we begin in \S\ref{s:general} with a discussion of the general behaviour of the entanglement entropy $\SA$ as function of the size parameter $\alpha$ and argue that in the holographic context, $\SA(\alpha) $ and consequentially $\dS(\alpha)$ should be continuous functions. We then proceed to see the explicit behaviour of entanglement entropy in 1+1 dimensional CFTs on a torus in \S\ref{s:s2d} and in higher dimensional CFTs in \S\ref{s:shd}. After displaying explicit examples of the entanglement plateaux, we  outline the connection with causal wedges in \S\ref{s:causal}.  We then step back to compare the prescriptions involving minimal (RT) versus extremal (HRT) surfaces and relatedly the role of the homology requirement in \sec{s:homol} and conclude with a discussion of open issues in \S\ref{s:discuss}.  Some technical aspects of finding the minimal surfaces are relegated to the Appendices.
\section{Generic behaviour of holographic $\SA(\alpha)$}
\label{s:general}

We begin our discussion by explaining the continuity of $\SA(\alpha)$.
Importantly, we focus on the RT prescription for static configurations, where the entire problem can be formulated on a $d$ dimensional Riemannian bulk geometry with $(d-1)$-dimensional boundary (in the conformal class of) $\Sigma$.
Consider a family of boundary regions $\ra_\alpha  \subseteq \Sigma$ specified by a real number $\alpha$.  For example, we can fix the shape of $\ra$ and let $\alpha$ denote the overall size; in the simplest case of $(d-2)$-spherically symmetric $\ra$ on a spatial slice of ESU$_d$ boundary spacetime, 
we take $\alpha \in [0,1]$ to be the fractional volume of the system.\footnote{
With the standard $SO(d-1)$ symmetric metric on spatial sections of ESU$_d$, i.e., $ds_\Sigma^2 = d\ph^2 + \sin^2\ph \, d\Omega_{d-2}^2$, we have 
$\alpha = \int_0^{\phA}(\sin\theta)^{d-2} d\theta /\int_0^\pi (\sin\theta)^{d-2} d\theta = \frac{1}{2} - \frac{\Gamma(\frac{d}{2})}{\sqrt{\pi}\, \Gamma(\frac{d-1}{2})}\; \cos\phA\; _2F_1\left(\frac{1}{2},\frac{3-d}{2},\frac{3}{2}, \cos^2\phA\right)$ for a polar cap characterized by co-latitude $\phA$.}

Now, consider the function $S(\alpha)$ defined by a {\it minimization} of area over all smooth bulk surfaces homologous to $\ra_\alpha$.  Starting from any surface homologous to $\ra_\alpha$, we may allow it to relax, continuously decreasing the area, such as by the mean curvature flow described in Appendix \ref{s:flows}. Since the area is bounded from below, it must tend to a limit, of which one quarter in Planck units is given by $S_k(\alpha)$. This may take several values, labeled by $k$, depending on the initial surface chosen, corresponding to different local minima of the area functional. Generically, the surface itself will also converge to an endpoint\footnote{
For some special spacetimes, such as extremal black holes with an infinite throat, this need not be strictly true, though it does not materially affect the argument.}, which will be a corresponding minimal surface $\mins{\alpha}^k$. As $\alpha$ is varied, we expect these minima to come in a discrete set of families depending smoothly on some parameter (since $\ra_\alpha$ and the spacetime are smooth), giving a set of curves in the plane of $\alpha$ and area. This means we have a set of functions $S_k(\alpha)$, defined on some interval of $\alpha$, continuous, and smooth away from critical points of $\alpha$, with $S(\alpha) = \min_k S_k(\alpha)$.

We now wish to consider the behaviour of $S(\alpha) \equiv \min_k S_k(\alpha)$. In simple spacetimes, such as pure AdS and small deformations thereof, there is only a single family $k=1$ which covers the full range of $\alpha$, so in such cases $S(\alpha)$ is manifestly smooth. However, it may happen at some point that two families, $k=1,2$, exchange dominance such that $S(\alpha)=S_1(\alpha)<S_2(\alpha)$ for $\alpha$ smaller than some critical value $\alpha_\trans$, and $S(\alpha)=S_2(\alpha)<S_1(\alpha)$ for $\alpha > \alpha_\trans$.   At $\alpha = \alpha_\trans$, $S_1=S_2=S$, so $S(\alpha)$ is necessarily continuous.  However, there will generically be a discontinuity in $\partial_\alpha S(\alpha)$, so $S(\alpha)$ has a kink.

We argue that this discontinuity of the derivative is as bad as it can get, and $S(\alpha)$ itself can never be discontinuous. For suppose $S(\alpha)$ has a jump discontinuity at some value $\alpha = \alpha_m$, so $S(\alpha)$ has different limits $S_-<S_+$ from the left and right respectively. Then by taking some surface with $\alpha$ less than, but sufficiently close to, $\alpha_m$, and area sufficiently close to $S_-$, and deforming slightly, we expect to be able to find a surface with $\alpha>\alpha_m$, but area still less than $S_+$: the variation in $\alpha$ may be taken as small as desired to temper the change in area. This is in contradiction with $S_\alpha$ being the minimum of the area functional. If desired, this deformed surface can be considered as the initial surface of a mean curvature flow.  This argument precludes, for example, a discontinuity when new families of minimal surfaces appear at $\alpha = \alpha_m$, and do not exist for $\alpha < \alpha_m$. This creation can happen, and typically such families appear at $\alpha_m$ in pairs (we shall present an example in \fig{f:AresAdS} below), but when this happens the area of the new family must exceed the area of some existing family.

The above argument indicates that within the RT prescription, the entanglement entropy of a given region $\ra$ should vary continuously with the parameters specifying the geometrical attributes of $\ra$.  In the following two sections, we will see this behaviour realized manifestly, even in situations where new families of minimal surfaces get nucleated at some intermediate $\alpha$.  By examining these families more closely, we will identify examples with large multiplicities of minimal surfaces.

\section{Entanglement entropy in $1+1$ dimensional CFTs}
\label{s:s2d}

In the previous section we have argued that for the holographic entanglement entropy given by a minimization procedure, $\SA(\alpha)$ must be continuous but need not be differentiable.  The lack of differentiability is typically associated with two families of minimal surfaces exchanging dominance.  Such an occurrence is not new, and good examples already exist in the literature.  The simplest one occurs for the bulk spacetime being the BTZ black hole, where this point was appreciated already in the early days \cite{Ryu:2006bv, Headrick:2007km, Azeyanagi:2007bj} and fleshed out a bit more explicitly in \cite{Blanco:2013joa}. We quickly review this story to illustrate the contrast with our other examples in higher dimensions.

\begin{figure}
\begin{center}
\includegraphics[width=2.8in]{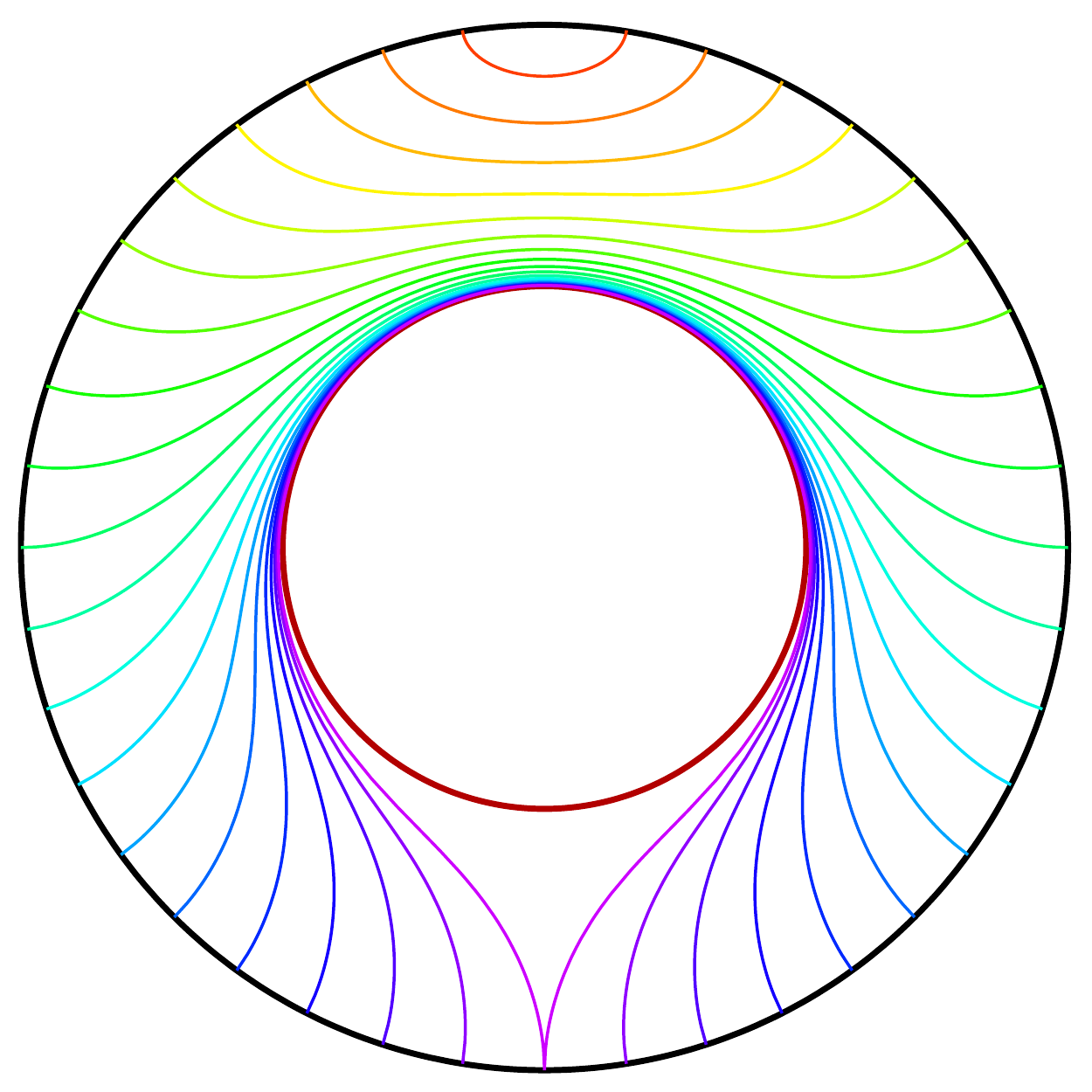} 
\hspace{.4cm}
\includegraphics[width=2.8in]{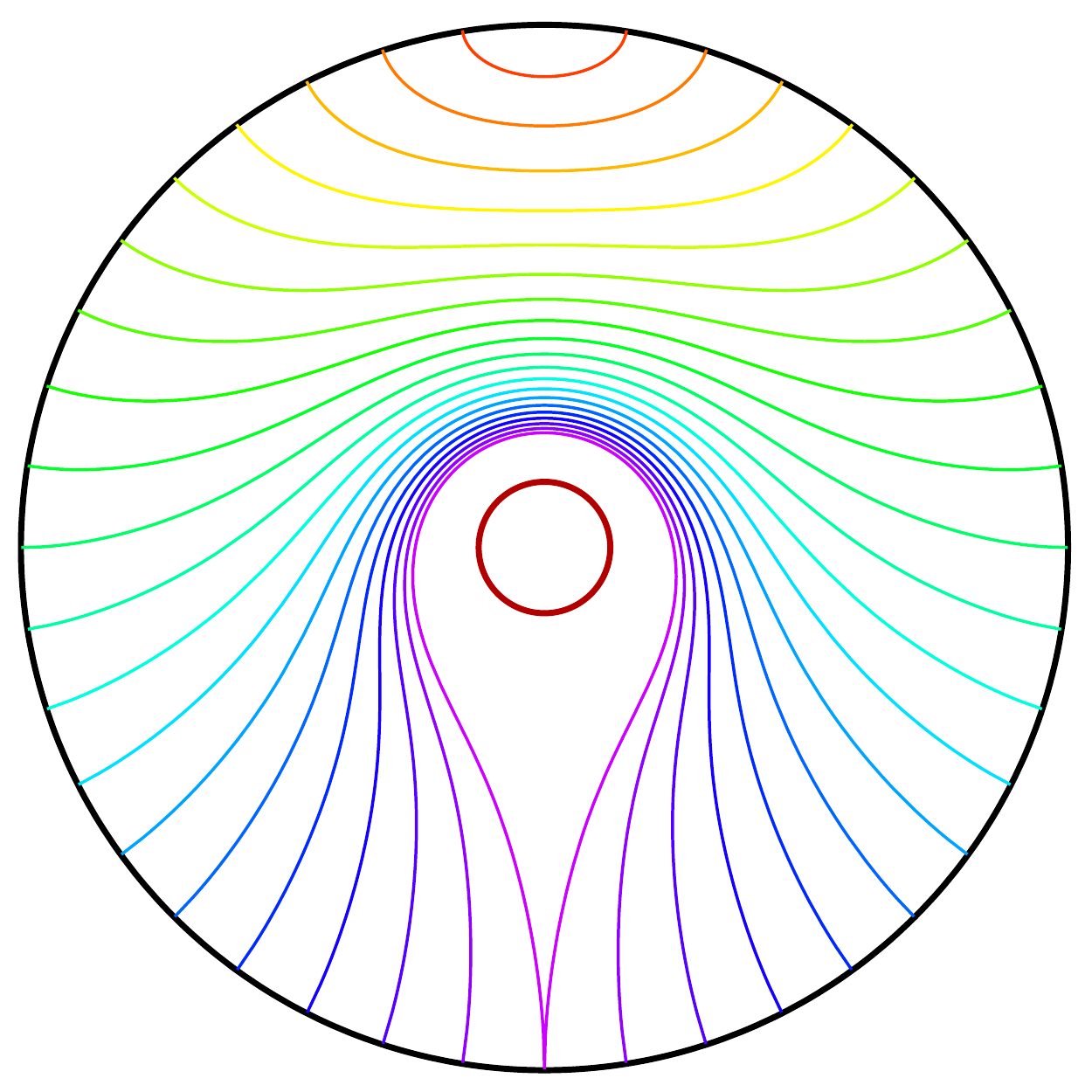} 
\caption{
Minimal surfaces (geodesics) in BTZ geometry with black hole size $\rh=1$ (left) and  $\rh=0.2$ (right). 
In each panel, the thick black circle represents the AdS boundary, and the thick red circle the black hole horizon.  The minimal surfaces are depicted by the thin curves, color-coded by  $\phA$, with $\phA$ varying between $0$ (red) and $\pi$ (purple) in increments of $0.05\pi$.
For ease of visualization we use a compactified radial coordinate $\rho = \tan^{-1} r$.  The boundary region $\ra$ is centered at $\ph = 0$ which is plotted on top.  
}
\label{f:BTZgeods}
\end{center}
\end{figure}

The metric for the BTZ black hole is given by\footnote{
We work in units where the AdS length $\ell_\text{AdS} =1$ and also set the radius of the boundary circle $R$ parameterized by $\ph \in [0,2\pi]$ to unity. It is easy to restore dimensions when necessary as we illustrate later.}
\begin{equation}
ds^2 = -f(r)\, dt^2 + \frac{dr^2}{f(r)} + r^2 \, d\ph^2 \,, \qquad f(r) =  r^2-\rh^{2}  \ .
\label{BTZmet}
\end{equation}	
It is a simple matter to find the minimal surfaces for regions $\ra = \{\ph: |\ph| \le \phA\}$ since these are given by spacelike geodesics. The result is best described by writing down  the spatial projection of the surfaces \cite{Hubeny:2012wa} 
\begin{equation}	 
\mins{1}(\phA): \qquad \bigg\{(r,\ph):\; r = {\mathbf \gamma}(\ph,\phA,\rh) \equiv  
\rh  \, \left( 1 - \frac{\cosh^2 (\rh \, \ph)}{ \cosh^2 (\rh \, \phA)} \right)^{\! \! -\frac{1}{2}} \ \bigg\} 
\label{BTZmins}
\end{equation}	
and is plotted in \fig{f:BTZgeods} for large and small black holes, for a set of $\phA \in [0,\pi]$.

As can be easily seen in \fig{f:BTZgeods}, connected spacelike geodesics (satisfying the homology constraint) always exist for any $\phA$ and $\rh$.  This makes sense, since there is no reason for the geodesics to break up (in fact, spacelike geodesics can orbit the black hole arbitrarily many times before returning to the boundary, albeit at the expense of greater length).  As we will see in \sec{s:shd}, this is in stark contrast to the behaviour of co-dimension 2 surfaces in higher dimensional black hole spacetimes.  Also note that in the BTZ case, another effect of the low dimensionality is that arbitrarily small black hole ($\rh \to 0$ in \req{BTZmet}) always looks effectively large in terms of the effect it has on geodesics.

Let us now consider the proper length along these geodesics and compute the entanglement entropy. As is well known, this computation reproduces the CFT computation of Cardy-Calabrese \cite{Calabrese:2004eu} for thermal CFT on the infinite line.  Naive application of the RT result for the CFT on a cylinder leads to\footnote{
We use CFT central charge determined by the Brown-Henneaux analysis $c = \frac{3\, \ell_\text{AdS}}{2\, G_N^{(3)}}$ and note that $T = \frac{\rh}{2\pi\,\ell_\text{AdS}\,R}$ is the temperature of the thermal density matrix $\rho_\Sigma$.}
\begin{equation}
\left(\SA \right)_\text{naive} = \frac{c}{3}\, \log \left(\frac{2\, r_\infty}{\rh} \, \sinh (\rh\,\phA) \right) \,, \qquad \forall\; \phA \in [0,\pi] 
\label{}
\end{equation}	
where $r_\infty$ is a radial cut-off to regulate the proper length of the geodesic and is our proxy for the boundary UV cut-off scale. It is then a simple matter to check that the naive result for $\dS$
\begin{equation}
\left(\dS\right)_\text{naive} = \frac{c}{3}\, \log \bigg[\sinh (\rh\,\phA)\,  \text{csch}(\rh \, (\pi-\phA) \bigg] 
\label{dsnaive}
\end{equation}	
violates the Araki-Lieb inequality. For the BTZ black hole $\Srho = \frac{\pi\,\rh}{2\, G_N^{(3)}} = \frac{2\pi^2}{3}\,c\, T \, R$, which is clearly bounded, while \eqref{dsnaive} diverges as $\phA \to 0$.

\begin{figure}
\begin{center}
\includegraphics[width=2.8in]{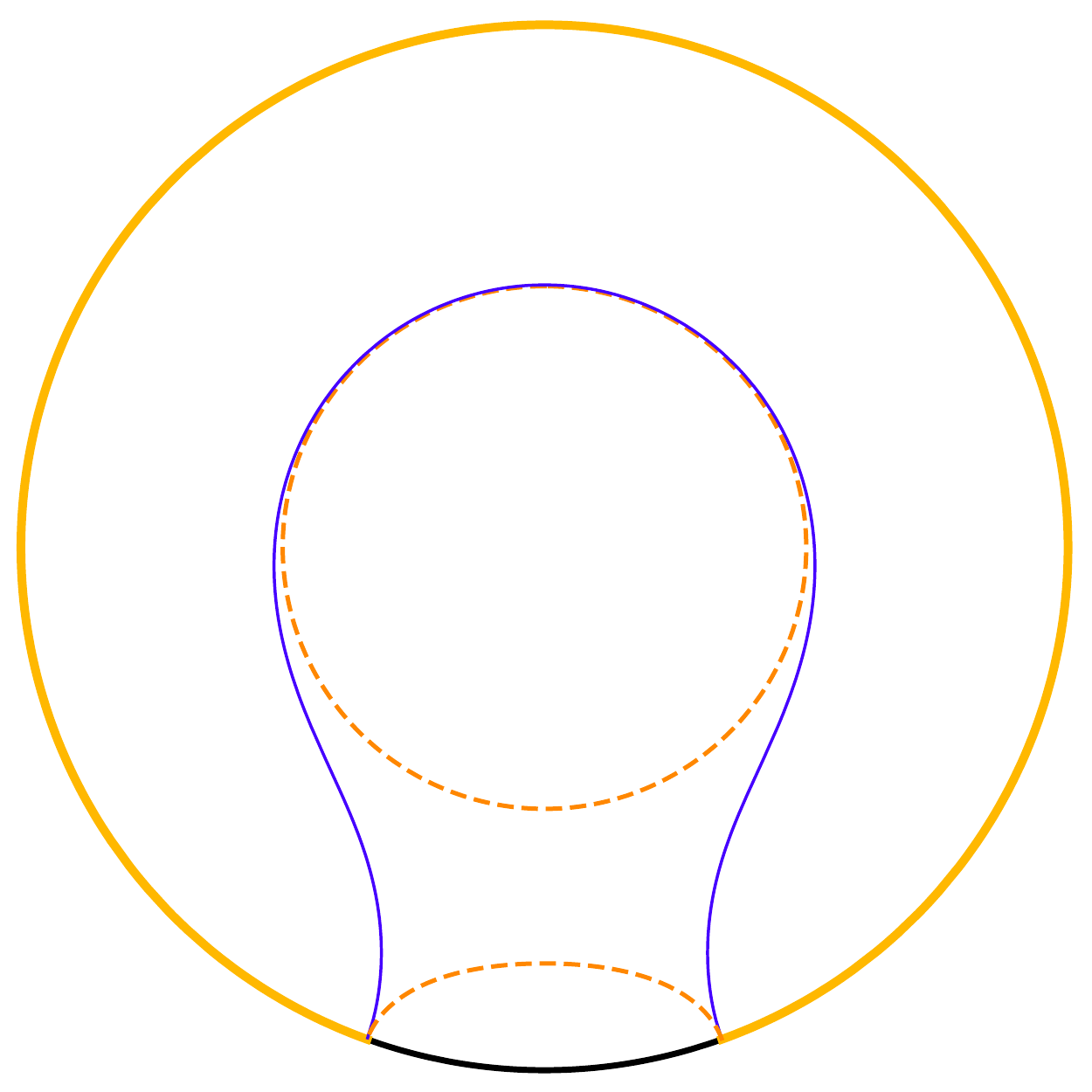} 
\hspace{.4cm}
\includegraphics[width=2.8in]{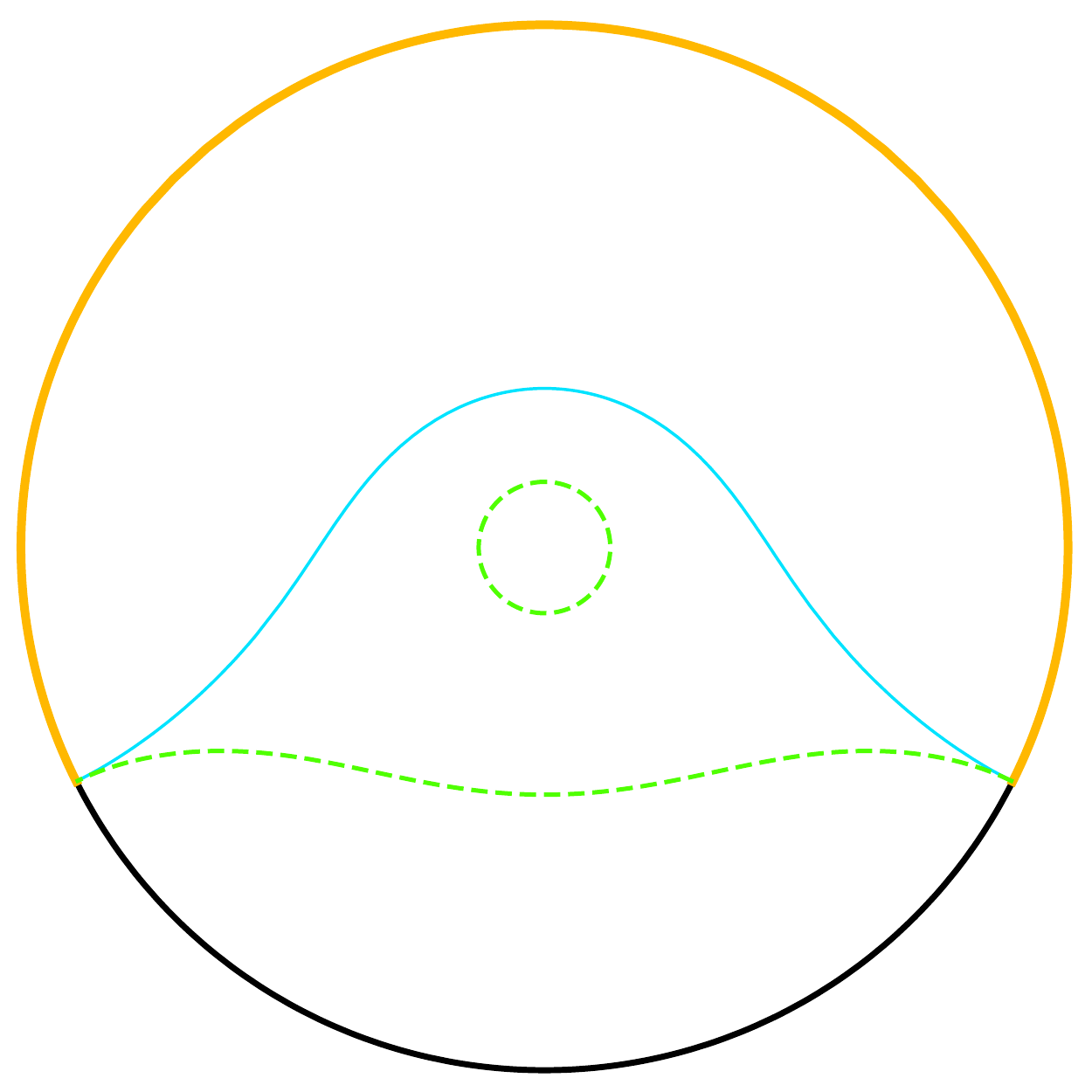} 
\caption{
Comparison of the two families $\mins{1}$ and $\mins{2}$ of the 
minimal surfaces (geodesics) in BTZ, with black hole sizes $\rh=1$ (left) and  $\rh=0.2$ (right)  as in \fig{f:BTZgeods}, plotted at the critical value of $\phA$ where they exchange dominance, i.e.\ at $\phA = \phA^\trans$.  The connected family $\mins{1}$ is represented by the solid curve (color-coded by $\phA$ as in \fig{f:BTZgeods}), while the disconnected family $\mins{2}$ is given by the two dashed curves (color-coded by $\pi-\phA$).
The thick orange arc on the boundary represents the region $\ra$.
}
\label{f:BTZtranss}
\end{center}
\end{figure}

The basic point is simply the following: for regions $\ra$ which are sufficiently large $\phA \ge \phA^\trans > \pi/2$ 
the holographic entanglement entropy is not computed from the connected minimal surface $\mins{1}$ but rather from a shorter two-component (disconnected) one $\mins{2}$.
Note that in any static black hole spacetime, the bifurcation surface is minimal by virtue of the null generator of the horizon vanishing there 
quadratically, forcing its extrinsic curvature to vanish. 
While  for any $\ra$ there exists a connected minimal surface which satisfies the homology condition, there also exists a disconnected minimal surface, given by the union of the (connected) minimal surface for $\rac$ and the bifurcation surface of the event horizon, which likewise does the job. 
Thus the second family of minimal surfaces, parameterized by $\phA$, is simply
 \begin{align}
\mins{2}(\phA):  \bigg\{ (r,\ph):\;r =  r_+ \;\; \cup \;\; r = {\mathbf \gamma}(\ph,\pi-\phA, \rh)\bigg\} \,,
 \label{}
 \end{align}
 where $ {\mathbf \gamma}$ is given by \req{BTZmins} (an example of both families of surfaces is plotted in \fig{f:BTZtranss} explained below).
Taking this fact into account we learn that the holographic entanglement entropy is given by 
\begin{align}
\SA(\phA) = \min \bigg\{ \frac{\text{Area}(\mins{1})}{4\, G_N^{(3)}}\;,\;\frac{\text{Area}(\mins{2})}{4\, G_N^{(3)}} \bigg\} \,,
\label{}
\end{align}
which then implies upon evaluating the lengths of the curves explicitly that 
\begin{equation}
\SA = 
	\begin{cases}
		&\frac{c}{3}\, \log \left(\frac{2\, r_\infty}{\rh} \, \sinh (\rh\,\phA) \right) \,,\hspace{3.5cm} \phA < \phA^\trans
		\\
		& \frac{c}{3}\,\pi\,\rh + \frac{c}{3}\, \log \left(\frac{2\, r_\infty}{\rh} \, \sinh (\rh\,(\pi - \phA)) \right) \,,\qquad \phA \geq \phA^\trans
		\end{cases} 
\label{salargec}
\end{equation}	
where we introduce the entanglement plateau scale $\phA^\trans$
to distinguish the dominant saddle point of the area functional. This can be explicitly evaluated as a function of the black hole size (or temperature) to be 
\begin{align}
\phA^\trans(\rh) = \frac{1}{\rh}\, \coth^{-1} \left(2\, \coth(\pi\, \rh) -1\right) .
\label{}
\end{align}
%
For orientation, we plot the two sets of minimal surfaces $\mins{1}(\alpha_\trans)$ and $\mins{2}(\alpha_\trans)$ at the transition value $\alpha_\trans =\phA^\trans/\pi$ in \fig{f:BTZtranss} (again for two black hole sizes for ease of comparison with \fig{f:BTZgeods}).  At this value of $\alpha$, $\mins{1}$ (solid curve) and $\mins{2}$ (dashed curves) have equal proper lengths.
Note that in the high temperature (large black hole) limit, the entanglement plateau scale approaches the size of the entire system:
\begin{equation}
\lim_{\rh \to \infty}\; \phA^\trans(\rh) = \pi \,.
\label{}
\end{equation}	

Translating this into field theory quantities and using $\alpha = \frac{\phA}{\pi}$ to parameterize the fraction of the system we consider, we have 
\begin{equation}
\alpha_{\cal X} = \frac{1}{2\pi^2\,T\,R} \, \coth^{-1} \left(2\, \coth\left(2\pi^2\, T\,R\right) -1\right)
\label{}
\end{equation}	
where $R$ is the radius of the boundary CFT cylinder. The main feature we want to illustrate is that for $\alpha > \alpha_\trans$
\begin{equation}
\dS = \Srho = \frac{2\pi^2}{3}\,c\, T \, R \qquad \Longrightarrow \qquad \SA = \SAc + \Srho
\label{ALsat}
\end{equation}	
as anticipated. Essentially for large enough boundary regions we can read off the thermal entropy by comparing directly the entanglement entropy of a region and its complement. The behaviour of $\dS$ for $1+1$ dimensional CFTs is shown explicitly later in \fig{f:dSnorm} (where we also demonstrate a similar feature in higher dimensional holographic field theories).

In deriving the relations \req{ALsat},
 we have implicitly assumed $\rh \geq 1$ (equivalently $T R \geq \frac{1}{2\pi}$), which is where the BTZ black hole is dual to the thermal density matrix for the CFT. For lower temperatures the density matrix is dual to thermal AdS geometry and the holographic computation described above should be modified. Geodesics in thermal AdS will actually give a result which says $\SA = \SAc$ at leading order in the $c\to \infty$ limit.\footnote{
 This is the statement that $\CO(1)$ corrections cannot be recovered from the classical gravity approximation.}
The thermal result will only be recovered by considering $1/c$ corrections. This makes sense since for $TR < \frac{1}{2\pi} $ one is in the `confined' phase of the CFT (the terminology is inherited from higher dimensions where one has an honest confinement/deconfinement transition in the large $N$ planar gauge theory). 

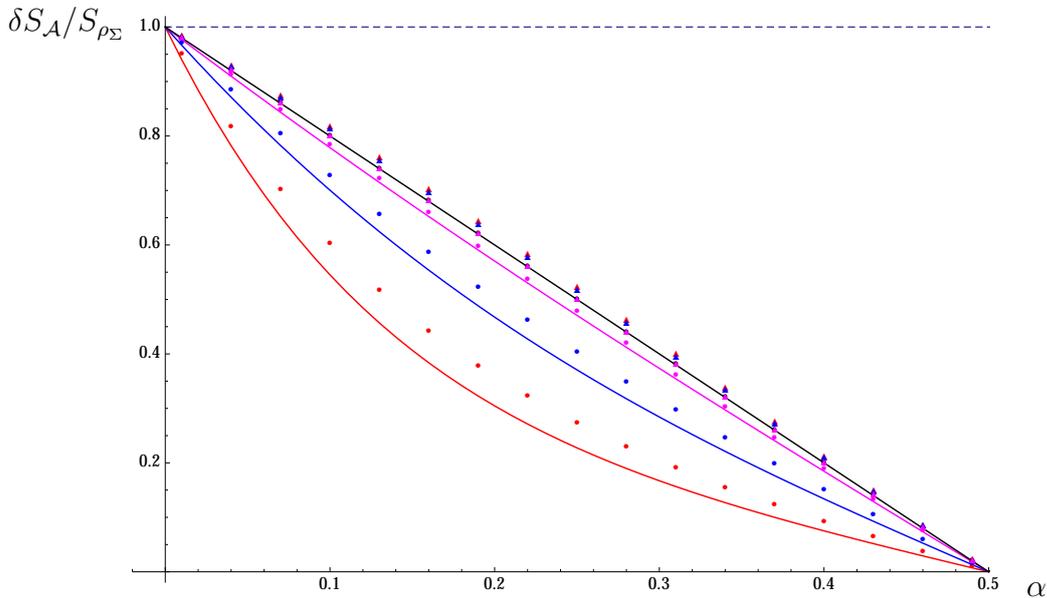
\begin{figure}
\begin{center}
\includegraphics[width=4.5in]{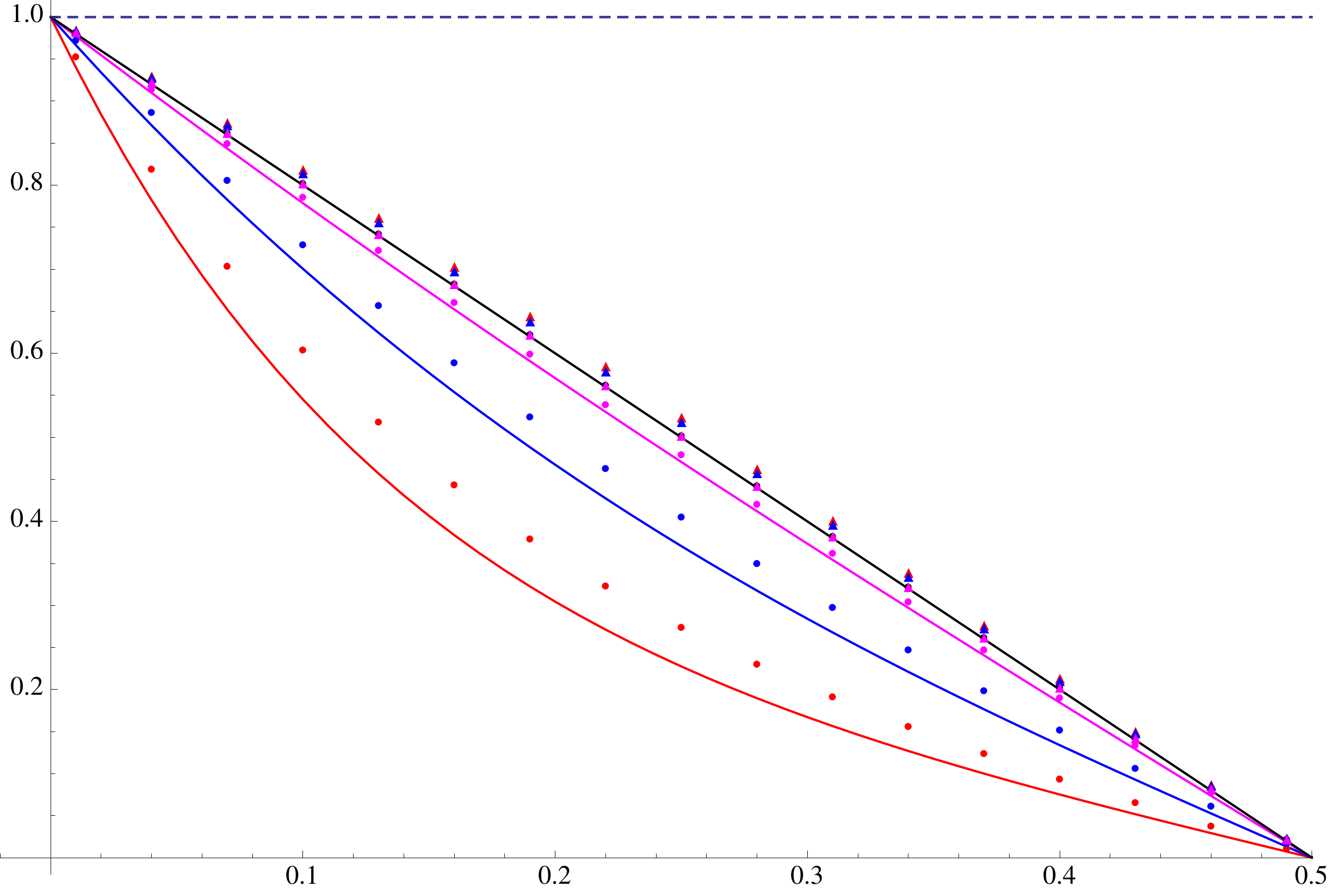} 
\begin{picture}(0,0)
\setlength{\unitlength}{1cm}
\put (-13.2,7.5) {$\dS/\Srho$}
\put(0.2,0){$\alpha$}
\end{picture}
\caption{
 Plot of the curves $\dS/\Srho$ for a Dirac fermion in $1 +1$ dimensions in the canonical ($T \neq 0 $) \cite{Azeyanagi:2007bj} and grand canonical ($T, \
\mu \neq 0 $) \cite{Ogawa:2011bz}. We examine the behaviour for a range of temperatures and chemical potential. The solid curves from bottom are $\beta =  4 $ (red), $\beta = 2 $ (blue), $\beta = 1 $ (magenta) and $\beta =  0.1 $ (black). 
The situation with the chemical potential turned on is indicated \
with the same colour coding with $\mu =  0.1 $ represented by circles and $\mu = 0.5 $ by triangles. The symmetry $\mu \leftrightarrow 1-\mu$ is used to restrict $\mu \in [0,1/2]$ and we see that for $\mu = 0.5$ the normalized $\dS$ is essentially the same at any temperature. This behaviour should be contrasted against the large $c$ holographic result displayed in \fig{f:dSnorm} for $d\geq 2$ thermal CFTs.}
\label{f:dS-Dirac}
\end{center}
\end{figure}

In $1+1$ dimensions we have also the ability to compare our large $c$ result for $\SA$ or $\dS$ with the behaviour encountered in small central charge systems, which is known for a couple of cases. The result for the entanglement entropy of a free neutral Dirac fermion $c=1$ was derived in \cite{Azeyanagi:2007bj} and was extended to a grand canonical density matrix including a $U(1)$ chemical potential in \cite{Ogawa:2011bz}.\footnote{
We note that angular momentum chemical potential was considered in \cite{Hubeny:2007xt}, which can be captured holographically using a rotating BTZ black hole in the large $c$ limit. However the ensemble is stationary (and not static) so one needs to work with extremal surfaces as emphasized there.} In either case we can use the resulting expressions (which being rather long we refrain from reproducing here) to understand the behaviour of the entanglement in the system. Using the explicit expressions one can check that the resulting answer exhibits no sharp feature. $\dS/\Srho$ is a smooth monotone function of $\alpha$, see \fig{f:dS-Dirac}. This is also consistent with the recent discussion of \cite{Herzog:2012bw} who compute analogous quantities in a harmonic chain which is a gapped theory and note that sharp features can only occur in the $c\to \infty$ limit (which as we mention earlier implements the thermodynamic limit).

Note that the Dirac fermion example is so far the only case for which the entanglement entropy of a thermal CFT in finite volume is explicitly known. In general one expects that  the details of the spectrum of the CFT play a crucial role in the entanglement, i.e., the answer is not a simple universal function of the central charge as the holographic result \eqref{salargec} suggests. It would be interesting to see if one can use the technology of \cite{Hartman:2013mia} to argue directly from a CFT analysis for an entanglement plateau relation like \eqref{ALsat} in the asymptotic large $c$ limit.

The behaviour of $\SA (\alpha)$, not surprisingly, is in fact rather similar to the behaviour of the entropy or free energy in the thermal ensemble (a point we will revisit in \S\ref{s:discuss}). In both cases there are multiple saddle points, which exchange dominance. For the thermal density matrix the
saddles are the thermal AdS geometry and the BTZ black hole both of which exist for the entire range of the dimensionless parameter $TR$. For the holographic entanglement entropy $\SA$ we have again two distinct saddles available for the entire range of parameters: the connected and disconnected surfaces $\mins{1}$ and $\mins{2}$ exist for all values of $\alpha \in [0,1]$.  For a sufficiently large region $\alpha \geq \alpha_\trans$, however, it is the disconnected surface that dominates and results in the entanglement plateau. We will soon see that existence of both families over the entire range of $\alpha$ is peculiar to three dimensions and the situation is much more intricate in higher dimensions.

\section{Entanglement entropy in higher dimensional field theories}
\label{s:shd}

In the previous section we saw that already in the 3-dimensional bulk spacetime corresponding to a thermal state in the dual CFT, we have multiple families of minimal surfaces $\mins{i}$ anchored at the boundary entangling surface $\entsurf$, for arbitrary $\ra$.  
Let us now examine the analogous situation in higher dimensions.  We will consider the \SAdS{d+1} bulk spacetime ($\ell_\text{AdS} =1$), 
\begin{equation}
ds^2 = -f(r)\, dt^2 + \frac{dr^2}{f(r)} + r^2 \, \left(d\ph^2 + \sin^2\ph\,  d\Omega_{d-2}^2\right) \,, \qquad f(r) =  r^2+1 - \frac{\rh^{d-2}  \,  (\rh^2 + 1)}{r^{d-2}} \ .
\label{SAdSdmet}
\end{equation}	
Large black holes $\rh \geq 1$ describe the thermal state on ESU (e.g.\ of ${\cal N}=4$ SYM 
on ${\bf S}^3 \times {\mathbb R}$ in the best-understood case of $d=4$), but we will consider black holes of any size for generality.  
For simplicity, we will take the boundary region $\ra$ to be a disk centered at the `north pole' $\ph=0$ with radius $\phA$, and consider only surfaces which maintain the residual $SO(d-1)$ spherical symmetry and remain at constant $t$.  This effectively reduces the problem to a 2-dimensional one: we can specify any such minimal surface as a curve in the $(r,\ph)$ plane.\footnote{
For plotting purposes, we will consider a compactified radial coordinate $\rho = \tan^{-1}  r$ as in \S\ref{s:s2d}, and double up the $\ph \in [0,\pi]$ coordinate to $\ph \in [-\pi,\pi]$ so that all curves will be reflection-symmetric.  The black hole will then be represented by a disk of radius $\tan^{-1} \rh$ centered at the origin.
}

The equations of motion for the minimal surface are obtained from the Nambu-Goto action for minimizing the area of the surface. Using an auxiliary parameter $s$ we have the Lagrangian for the system
\begin{equation}
{\cal L} = \left(r\, \sin\ph\right)^{d-2}\, \sqrt{\frac{1}{f(r)} \, \left(\frac{dr}{ds}\right)^2+ r^2\left(\frac{d\ph}{ds}\right)^2 } \,.
\label{lagmin}
\end{equation}	
The Euler-Lagrange equations for $r(s)$ and $\ph(s)$ are equivalent due to the reparameterization invariance, giving a second order ODE for $r(s)$ and $\ph(s)$. The second equation comes from choice of parameter.\footnote{
One might be tempted to use $\ph$ itself as a parameter, but it turns out that $\ph$ is not monotonic in higher dimensions. We choose $s$  such that the evaluation of the area integral on-shell reduces to 
$\int ds\, \left(r\, \sin\ph\right)^{d-2}$, for good behaviour numerically. In particular, the results for the regularized area of the surface are less error-prone since $r$ increases exponentially in $s$ as the boundary is approached.}

The topology of the problem constrains any surface to pass through a pole of the sphere, so to classify all connected surfaces we may start integration (w.l.o.g.) at the North pole. Requiring smoothness of $\mins{}$ there, we obtain a one-parameter family of minimal surfaces, specified by the `initial value' for the radial coordinate at the north pole, $r_0 \equiv r(\ph=0)$.\footnote{
Apart from ODE methods, the numerical construction of the surfaces is also done using a mean curvature flow. The reader interested in the details is encouraged to consult Appendix \ref{s:flows} where we outline the necessary mathematical  technology and the algorithm used for obtaining the surfaces.}
Given $r_0$, we find the minimal surface and from this read off the latitude $\phA$ which it is anchored on.  From this we will classify the boundary region size by $\alpha$, the proportion of the area contained in the boundary region homologous to the surface: $\alpha = \frac{\text{Vol}(\ra)}{\text{Vol}({\bf S^{d-1}})}$ which tends to $0,1$ as $\phA\to0,\pi$ respectively.

\subsection{Entanglement plateaux in $d>2$}
\label{s:entplatd}

Considering $\alpha$ as a function of initial radius $r_0$, the situation is rather different from the $d=2$ case. In $d>2$, $\alpha$ reaches a maximum value of $\alpha_m<1$: for sufficiently large regions, there are no connected minimal surfaces obeying the homology constraint! This can be intuited from analogous behaviour in the classic `soap bubble' problem in flat space, of finding a minimal surface between two circular rings. To reduce the area, there is a tendency to shrink the radius of the tube, counteracted by the constraint of ending on the rings. But when the ratio of ring radius to separation is sufficiently small, the rings do not hold the surface up enough to prevent the tube radius from shrinking to zero, and the surface separates into two disconnected parts. This reasoning carries over directly to our set-up, and the process of surfaces splitting into disconnected pieces can be seen explicitly in the animations of our numerical simulations of mean curvature flow, to be found on the arXiv as  \href{http://arxiv.org/src/1306.4004/anc}{ancillary files} for the submission.  The area cost of having a wide tube is greater in higher dimensions, which leads one to expect that $\alpha_m$ should be smaller, reflecting the greater tendency of the surfaces to split up.\footnote{
This behaviour is in fact reminiscent of similar observation in gravitational context of the Gregory-Laflamme instability of higher-dimensional black strings (or branes), which are likewise more prone to fragmenting with increasing dimension \cite{Caldarelli:2008mv,Emparan:2013moa}.
}
The physical point is that this leaves no option but to consider the disconnected surfaces, to which we now turn. In the next section, we offer a very different geometrical justification of why connected extremal surfaces homologous to $\ra$ cannot exist beyond a certain $\phA$, which is based on causality in the full Lorentzian geometry.

So far, we have a one-parameter family of connected surfaces $\mins{1}$. In addition to this, we must consider the `disconnected' family $\mins{2}$ of surfaces with two connected components: the first, anchored to the boundary, is a reflection of a connected surface already considered, passing through the South pole. Due to the nontrivial ${\bf S}^{d-1}\times\mathbb{R}$ topology of the static slice of the geometry, this alone fails to satisfy the homology constraint, so it is supplemented by a second piece, the bifurcation sphere on the event horizon of the black hole. To identify the correct entangling surface determining $\SA(\alpha)$, we must compare the areas of these two families; they exchange dominance at $\alpha_\trans<\alpha_m$, as is inevitable from continuity. This means that at this value $\alpha_\trans$, there are two surfaces of equal area, and as we vary $\alpha$ through $\alpha_\trans$, the surfaces jump and the entanglement entropy as a function of region size has a kink.
%
\begin{figure}
\begin{center}
\includegraphics[width=3in]{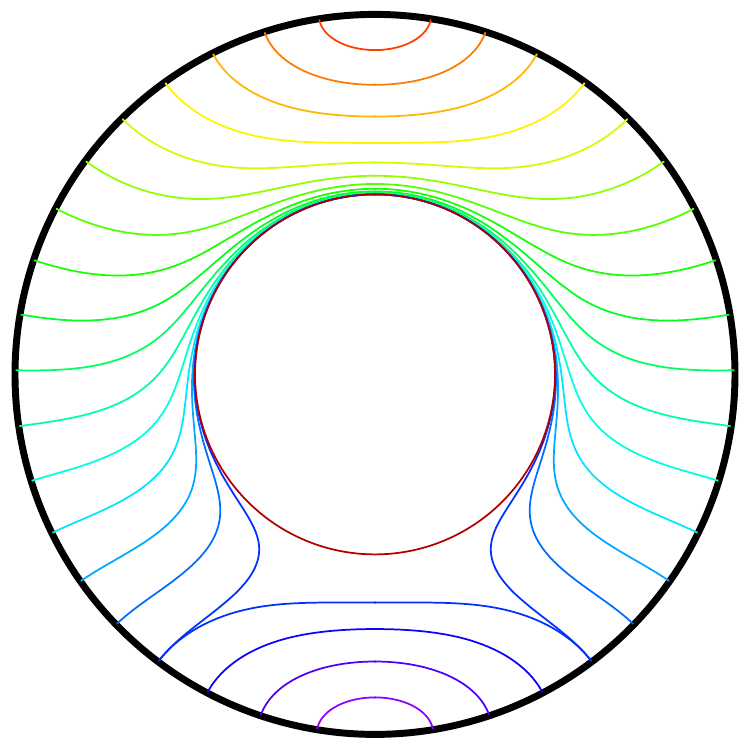} 
\hspace{.2cm}
\vspace{.4cm}
\includegraphics[width=3in]{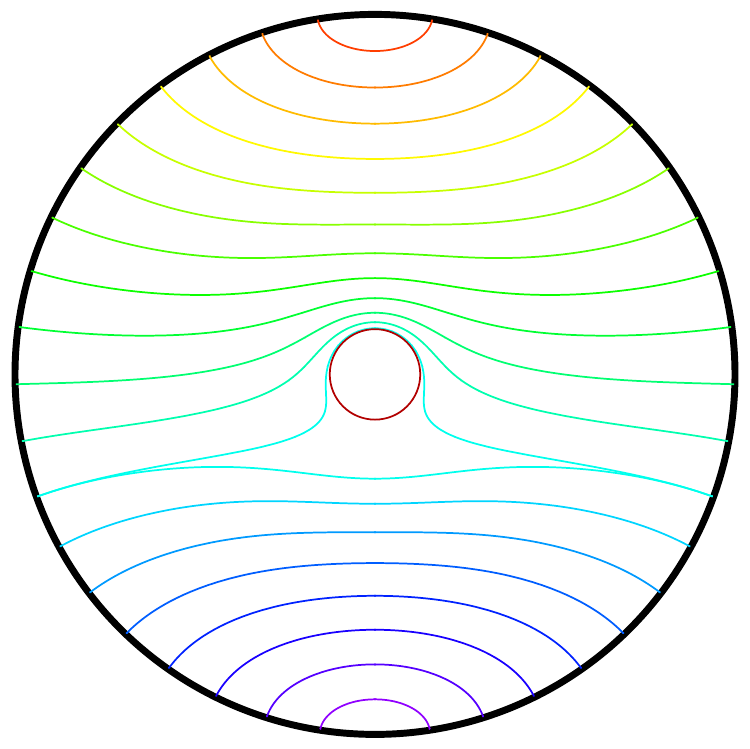} 
\includegraphics[width=3in]{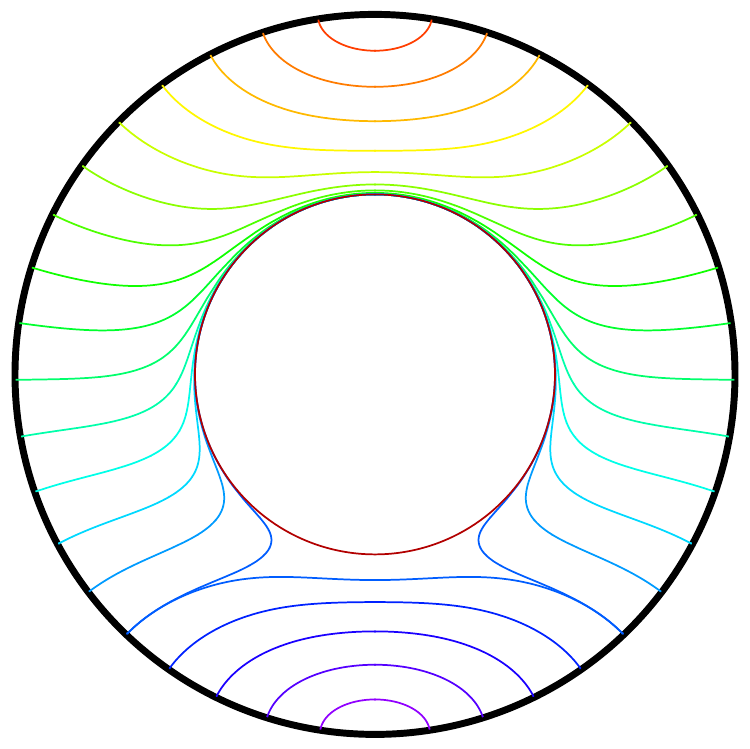} 
\hspace{.2cm}
\includegraphics[width=3in]{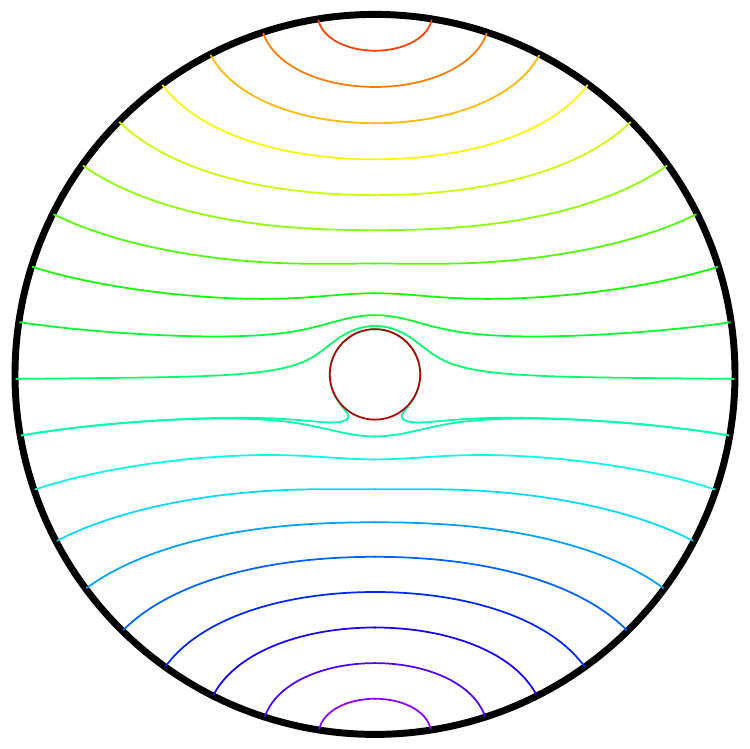} 
\caption{
Minimal surfaces for  $\rh=1$ (left) and  $\rh=0.2$ (right) black holes in \SAdS{} in $3+1$ (top) and $4+1$ (bottom) dimensions, analogous to the plots of \fig{f:BTZgeods}. 
In each panel, the thick black circle represents the AdS boundary, and the thin red circle the black hole horizon. The boundary region $\ra$ is centered at $\ph = 0$ (North Pole) which is plotted on top.  
 Further plots with other black hole sizes and higher dimensions can  be found as \href{http://arxiv.org/src/1306.4004/anc}{ancillary files} with the submission.}
\label{f:SurfacesAdS}
\end{center}
\end{figure}
In \fig{f:SurfacesAdS} we show the surfaces which determine $\SA(\alpha)$ for various black hole sizes and dimensions. Those passing `above' the black hole belong to the family $\mins{1}$, and those passing `below' should be supplemented by the horizon, and belong to  $\mins{2}$.
Comparing the panels horizontally, we see that the smaller the black hole, the smaller $\alpha_\trans$ gets; indeed, as $\rh \to 0$, $\alpha_\trans \to 1/2$, whereas as  $\rh \to \infty$, $\alpha_\trans \to 1$;  we plot the actual curve $\alpha_\trans(\rh)$ in \sec{s:causal}.
On the other hand, comparing the panels vertically, we see that as $d$ increases, the surfaces get less affected by the black hole until close to the horizon.  This is easy to understand from the simple fact that gravity falls off faster in higher dimensions, and chimes perfectly with the intuition recently explained in \cite{Emparan:2013moa}.
Turning to common characteristics, one universal feature of all cases is that none of the minimal surfaces (anchored on the boundary) can penetrate past the event horizon.  This is in fact true for any static geometry with a horizon  \cite{Hubeny:2012ry}.\footnote{
More specifically,  \cite{Hubeny:2012ry} showed that for asymptotically AdS spacetimes with planar symmetry the deepest reach (i.e.\ the turning point, furthest away from the boundary) of any extremal surface which is fully anchored on (a single) AdS boundary cannot occur inside a black hole.  On the other hand, in time-dependent spacetimes, there do exist extremal surfaces anchored on the boundary which penetrate past the horizon \cite{Hubeny:2012ry, AbajoArrastia:2010yt,Hartman:2013qma, Liu:2013iza, Hubeny:2013uq, Shenker:2013pqa}.
For this reason, entanglement entropy is cognizant of some physics inside the horizon. 
}

It is evident from the symmetry of our setup that in the regime $\alpha \ge \alpha_\trans$, we have the identity
\begin{equation}
\SA = \SAc + \Srho
\label{ALsaturd}
\end{equation}	
so that we manifestly saturate the Araki-Lieb inequality. This is identical to the situation described in \S\ref{s:s2d}. Indeed, a plot of $\dS$ as a function of $\alpha$ for various dimensions explicitly reveals the entanglement plateaux as illustrated in \fig{f:dSnorm} where we include the result for $d=2$ CFTs as well for completeness.
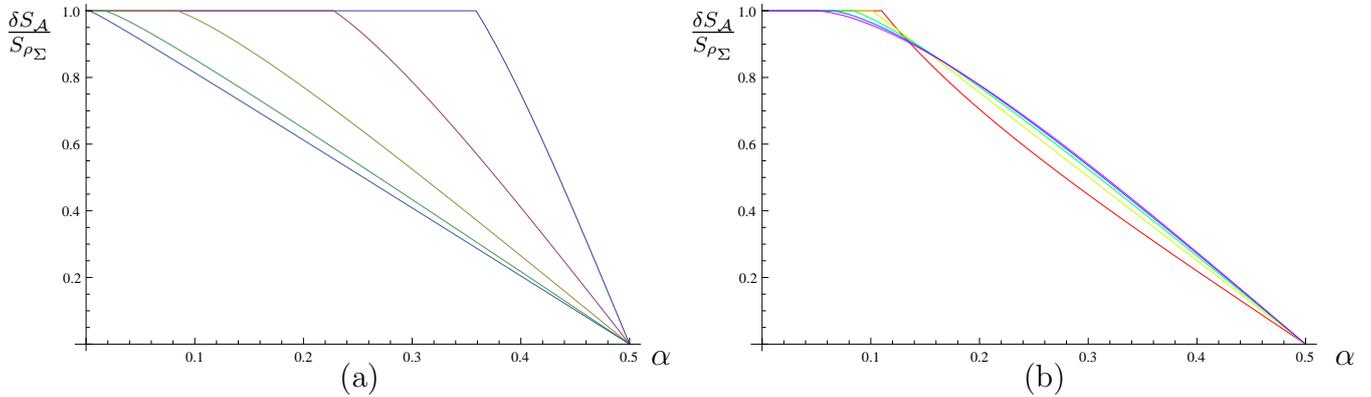
\begin{figure}
\begin{center}
\includegraphics[width=3in]{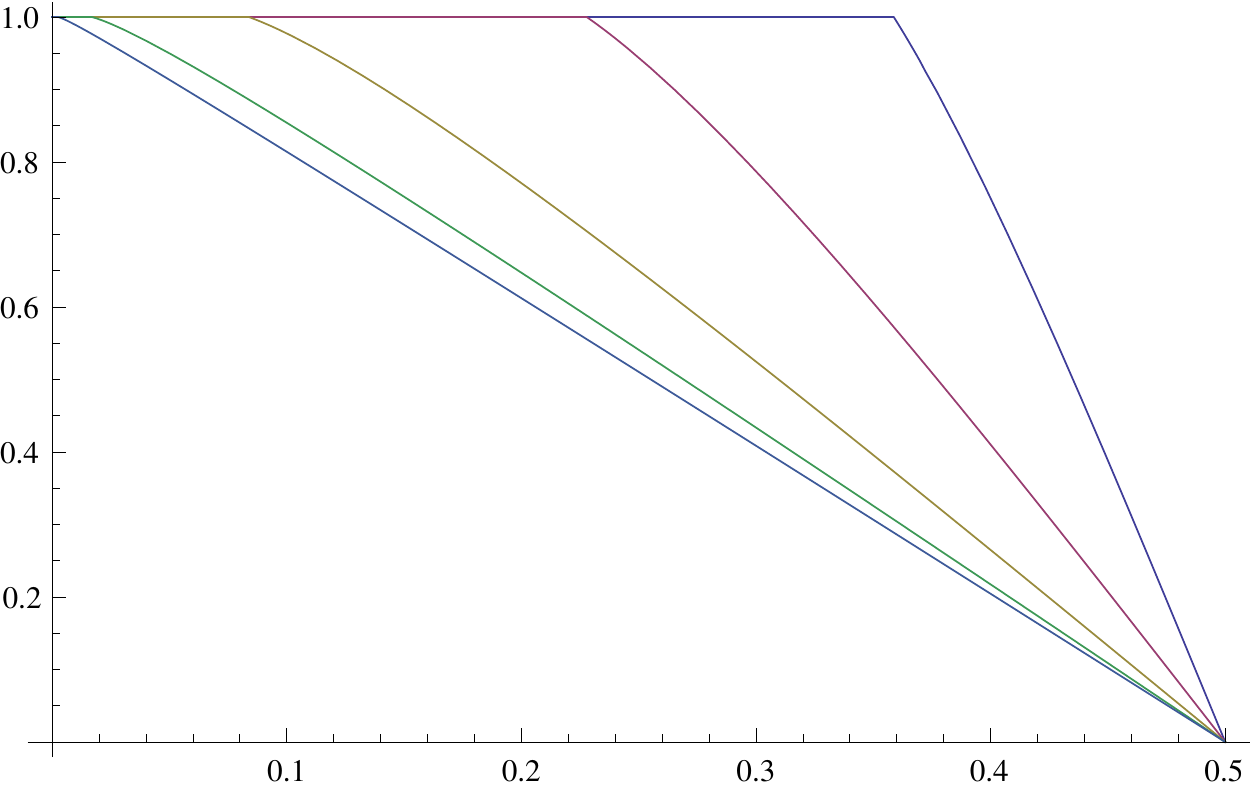} 
\hspace{1cm}
\includegraphics[width=3in]{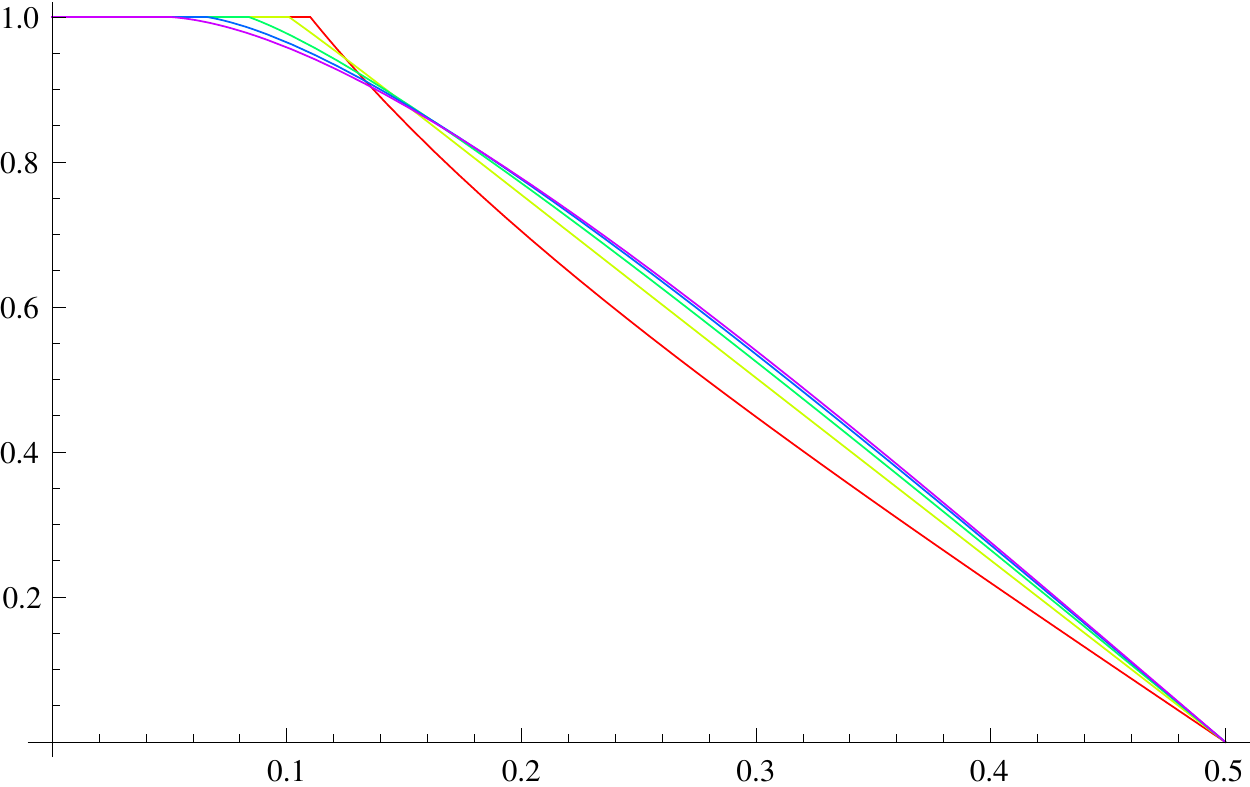} 
\begin{picture}(0,0)
\setlength{\unitlength}{1cm}
\put (-17.4,4.3) {$\frac{\dS}{\Srho}$}
\put (-8.4,4.3) {$\frac{\dS}{\Srho}$}
\put(0.1,0){$\alpha$}
\put(-8.9,0){$\alpha$}
\put(-4,-0.3){(b)}
\put(-13,-0.3){(a)}
\end{picture}
\caption{
Entanglement entropy plateaux for \SAdS{} black holes dual to thermal field theories. (a).  Behaviour in \SAdS{}$_5$ for varying $\rh$. From right to left we have $\rh = \frac{1}{4}$ (blue), $\rh = \frac{1}{2}$ (purple), $\rh = 1$ (yellow), $\rh = 2$ (green) and $\rh =4$ (blue).  (b). Dimension dependence of the plateaux, at the Hawking-Page transition point $R \, T=\frac{d-1}{2\pi}$
for $d=2,3,4,5,6$ (red, yellow, green, blue, purple respectively). Note that  the horizon size is related to the temperature via $\rh=\frac{2\pi}{d} \left(R\,T+\sqrt{(RT)^2-\frac{d(d-2)}{4\pi^2}}\right)$. }
\label{f:dSnorm}
\end{center}
\end{figure}
In \fig{f:dSnorm}(a) we show the behaviour for a fixed dimension $d=4$ as we vary the temperature by changing the black hole size, whereas in \fig{f:dSnorm}(b) we display the behaviour for different dimensions at fixed $TR = \frac{d-1}{2\pi}$.
We see that the entanglement plateau ends by a kink (e.g.\ in $d=2$ the behaviour of $\dS/\Srho$ can be obtained from \eqref{salargec}), the extent of which gets smaller with increasing dimension. To be sure, the precise details depend on the nature of the comparison; we have found it reasonable to fix the temperature (in units of the boundary sphere) and compare the behaviour across dimensions. In general the plateau width decreases as a function of $T$ for a given dimension (the behaviour as a function of $d$ depends on the precise value of $RT$; see \href{http://arxiv.org/src/1306.4004/anc}{ancillary files} for $RT = \frac{2}{3}$).

\subsection{Sub-dominant saddles: folds in minimal surfaces}
\label{s:}

Although we now have the curve  $\SA(\alpha)$ for the full range of $\alpha$, it is interesting to ask what happens to the connected minimal surfaces (and hence the disconnected ones as well, by the reflection symmetry) in their sub-dominant regime.  We know that a continuation of these families must exist, since the deepest radius $r_0$ did not cover the full range down to $\rh$.

As we decrease $r_0$ from the critical value $r_0(\alpha_\trans)$, the endpoint of the surface $\alpha$ increases beyond $\alpha_\trans$ as expected, but as previously noted, reaches a maximum at $\alpha = \alpha_m$,  beyond which there simply are no connected surfaces at all.  Instead, as one decreases $r_0$ further, the endpoints $\alpha$ start to recede to lower values again, even though the `neck' of the surface $\mins{1}$ near $\ph=\pi$ keeps closing off.
\begin{figure}
\begin{center}
\includegraphics[width=4in]{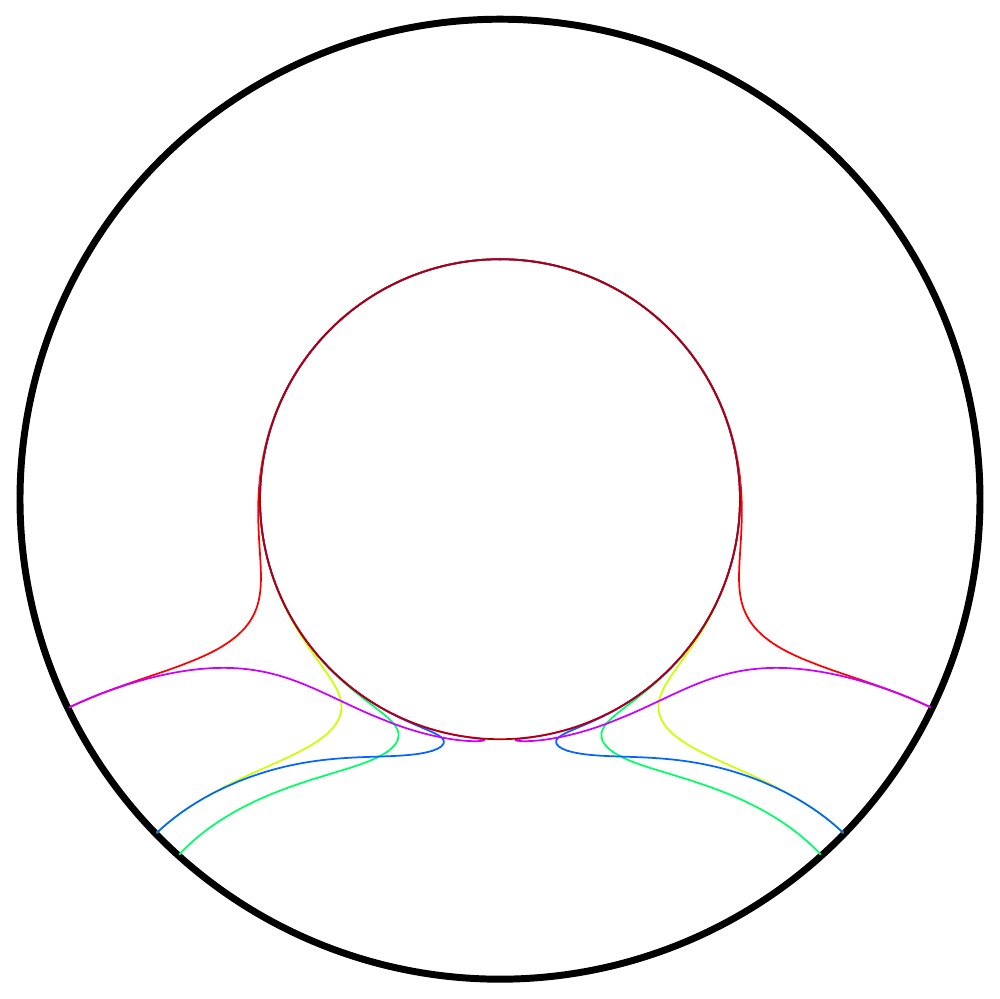} 
\caption{
Continuation of family $\mins{1}$ of minimal surfaces in \SAdS{5} for  $\rh=1$. As the turning point $r_0$ approaches the horizon, we see an intricate pattern of the minimal surfaces folding back onto themselves; solutions with $n$-folds are denoted $\mins{1,(n)}$.
We illustrate $\mins{1,(0)}$ (red, yellow) and $\mins{1,(1)}$ (blue, purple) surfaces for two specific choices of $\alpha$; these two families terminate at $\alpha = \alpha_m$ (green).}
\label{f:staticSurfaces}
\end{center}
\end{figure}
This behaviour is illustrated in \fig{f:staticSurfaces}.
Bringing $r_0$ still closer to the horizon reveals another surprise: the pattern repeats itself.  Eventually $\alpha$ reaches a minimal value (which happens to be very close to $1-\alpha_m$) and turns around again.  In other words, now the corresponding minimal surface has two necks, one near $\ph=\pi$ and the other near $\ph = 0$.

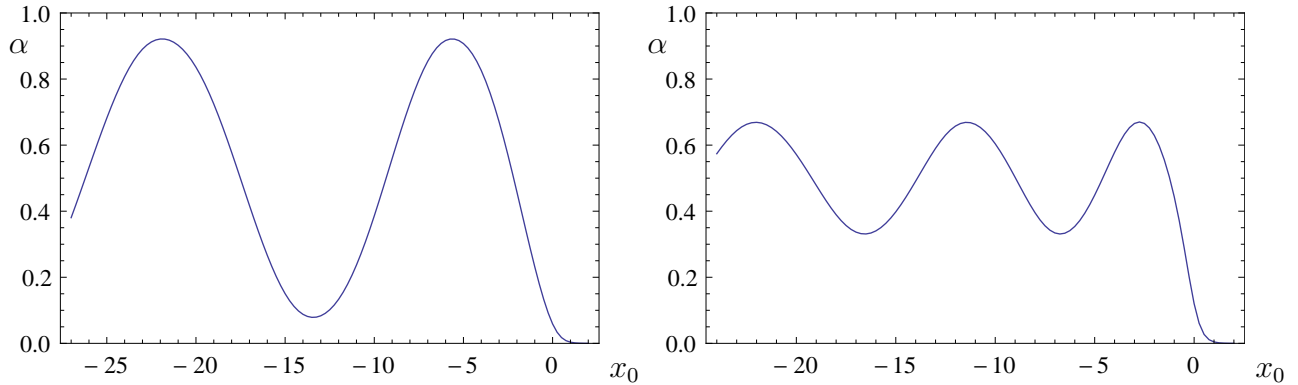
\begin{figure}[t]
\begin{center}
\includegraphics[width=3in]{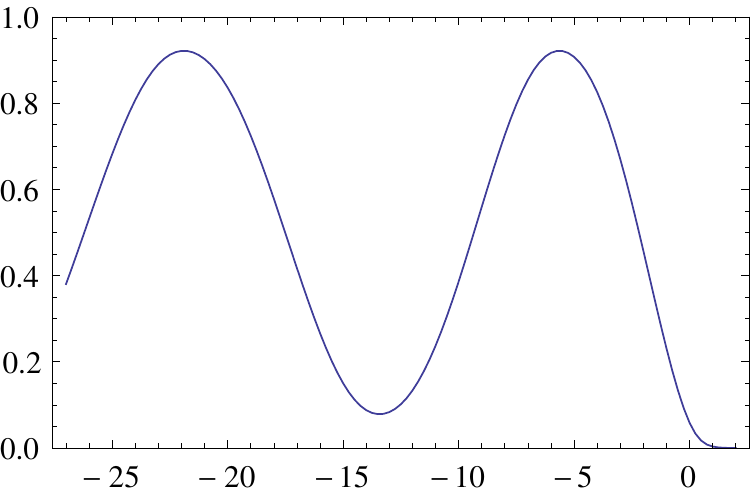} 
\hspace{.6cm}
\includegraphics[width=3in]{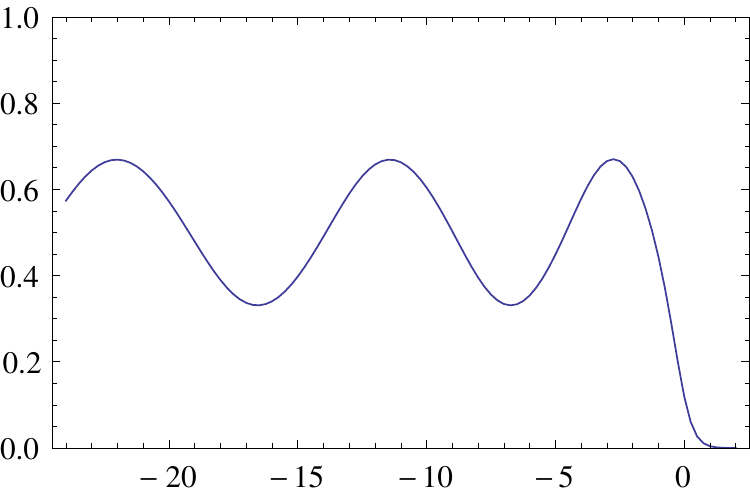}
\begin{picture}(0,0)
\setlength{\unitlength}{1cm}
\put (0,0) {$x_0$}
\put(-8,4.3){$\alpha$}
\put (-8.5,0) {$x_0$}
\put(-16.4,4.3){$\alpha$}
\end{picture}
\caption{
Near-horizon analysis of folds in the minimal surfaces for  $\rh=1$ (left) and  $\rh=0.2$ (right) black holes in \SAdS{4}. Here $x_0 = \frac{1}{2} \log(r_0 -\rh)$ measures the closest approach of the surface to the horizon.
}
\label{f:x0ThetaInfinityAdS4}
\end{center}
\end{figure}

Careful examination reveals that the closer the minimal radius $r_0$ is to the horizon, the more intricate the surface. The endpoints $\phA$ are restricted to lie sufficiently close to the equator $\ph = \frac{\pi}{2}$, in particular, $1-\alpha_m \lesssim\alpha\lesssim\alpha_m$,
as manifested in \fig{f:x0ThetaInfinityAdS4}, where we plot $\alpha$ as a function of rescaled coordinate $x_0 = \frac{1}{2}\, \log (r_0-\rh)$ indicating the proximity of $r_0$ to the horizon (cf Appendix \ref{s:nhorizon}). 

In fact, the pattern of minimal surfaces exhibits a discrete self-similarity. The minimal surfaces wrap the horizon multiple times, the distance from the horizon of each wrapping parametrically separated from the next. We denote the solution with $n$-folds 
 around the horizon as $\mins{1,(n)}$. The connected surfaces described earlier in \fig{f:SurfacesAdS} are $\mins{1,(0)}$ with this refinement in the notation. The existence of these solutions and their properties can be analytically understood using a local analysis around the turning points, and near the horizon away from the turning points, as we illustrate in Appendix \ref{s:nhorizon}. Further, this analysis is generic for any spherically symmetric non-extremal horizon, so there is nothing particularly special about \SAdS{} in this regard.

\begin{figure}[h!]
\begin{center}
\includegraphics[width=3.2in]{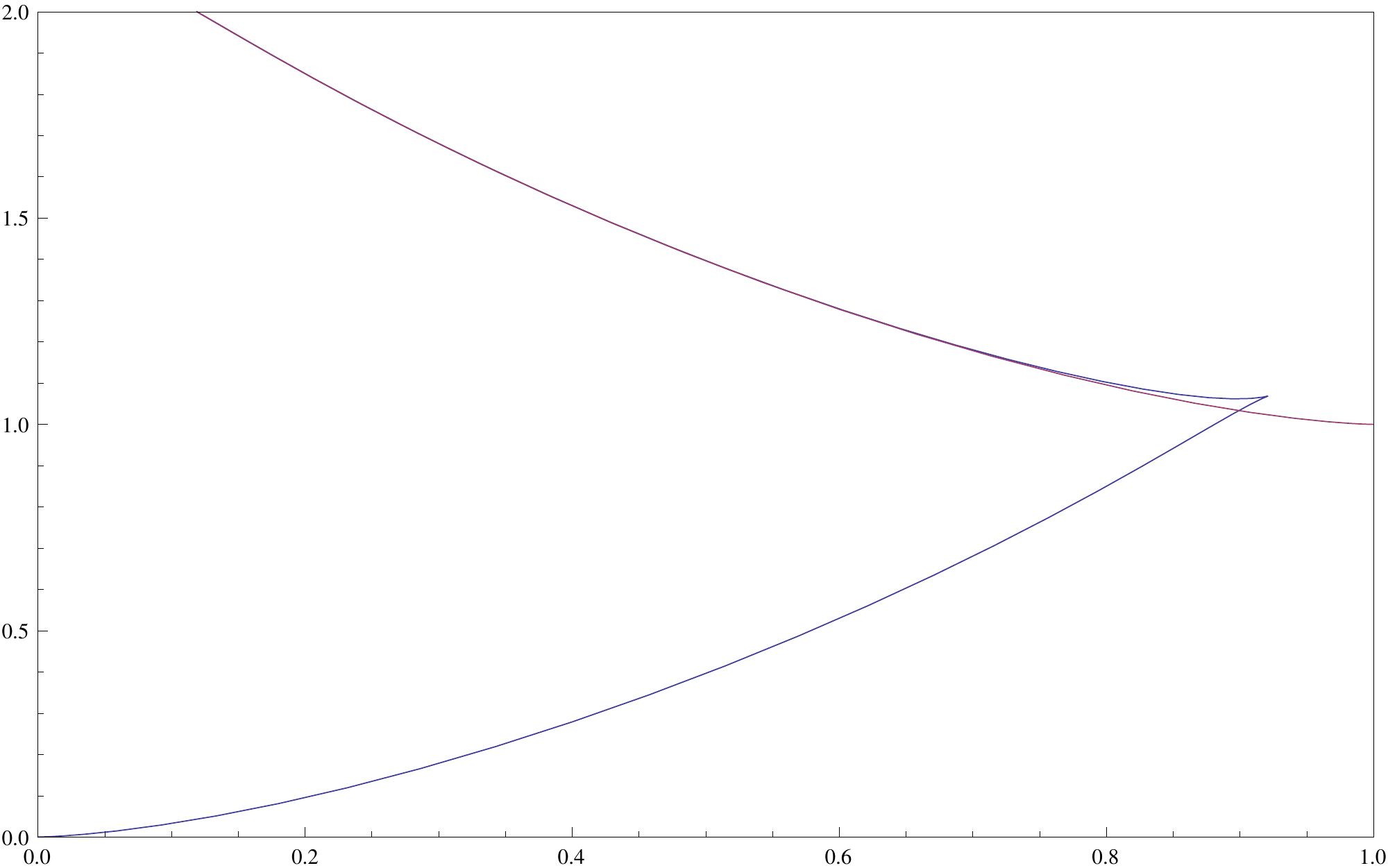} 
\hspace{1cm}
\includegraphics[width=3.2in]{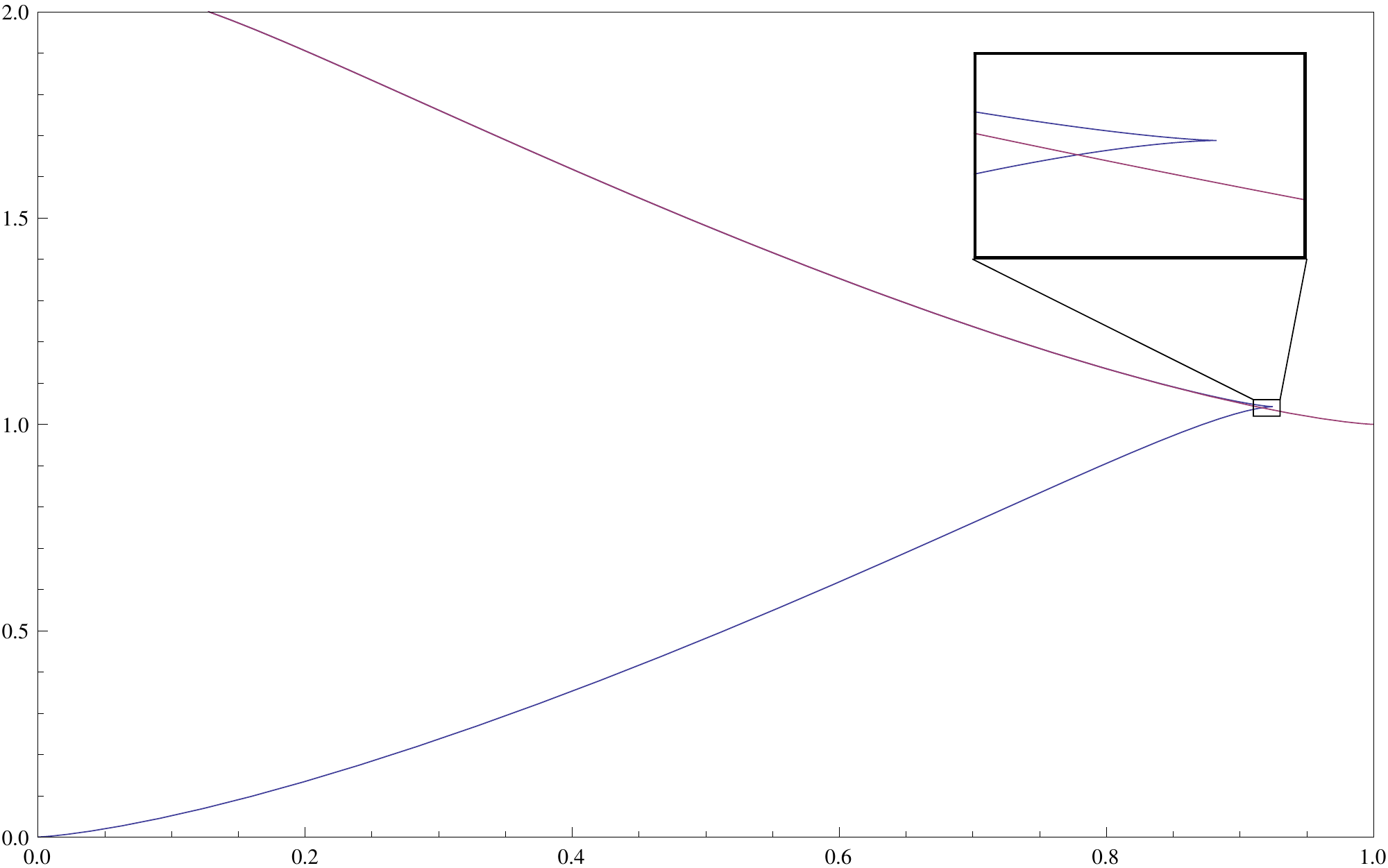}
\begin{picture}(0,0)
\setlength{\unitlength}{1cm}
\put (4.5,0)  {(b)}
\put (-1,0.2)  {$\alpha$}
\put(-9.4,5){$\frac{\text{A}({\mins{}})}{\text{A}_{BH}}$}
\put(-5,0){(a)}
\put (8.8,0.2) {$\alpha$}
\put(0,5){$\frac{\text{A}({\mins{}})}{\text{A}_{BH}}$}
\end{picture}
\caption{Regulated  of minimal surfaces (normalized in units of black hole area), for (a) \SAdS{4} ($\rh=1$), and (b) \SAdS{5} ($\rh=1$).  The blue curve is the family $\mins{1}$, and the purple curve the disconnected family $\mins{2}$. The area of the connected minimal surfaces grows without bound, since we encounter multiply-folded surfaces $\mins{1,(n)}$ which for $n>1$ which wrap the horizon multiple times. The swallow-tail behaviour characteristic of first order phase transitions is clearly visible in the plots. 
}
\label{f:AresAdS}
\end{center}
\end{figure}

Of course, such surfaces which fold around the black hole multiple times have larger area (approximately by the black hole area times the number of foldings) than their simpler cousins, so these surfaces are not directly relevant for the entanglement entropy.  This is already visible explicitly in \fig{f:AresAdS}, where we plot the actual areas of the minimal surfaces, as a function of $\alpha$. For $\alpha > \alpha_\trans$ the disconnected family $\mins{2}$ takes over, a fact we have already used to demonstrate the entanglement plateau in \fig{f:dSnorm}. In the regime $\alpha > \alpha_\trans$ the connected surfaces have greater area, and indeed the area continues to grow as the surface folds over about the horizon. It is intriguing to note that the difference $\text{Area}(\mins{1,(n)})- \text{Area}(\mins{1,(n-2)}) \approx 2\, \Srho$ 
increasingly more accurately as $n$ gets large. While the sub-dominant connected surfaces $\mins{1,(n)}$ for  discussed above are irrelevant for the boundary entanglement for $\alpha > \alpha_\trans$ it is 
nevertheless curious that such a simple geometry as \SAdS{} allows for such a rich structure of minimal surfaces!

%

\section{Causality and holographic entanglement}
\label{s:causal}

In \sec{s:shd} we have examined the families of minimal surfaces in static thermal density matrices for holographic theories which compute the entanglement entropy.
The main surprise was that beyond a certain size of $\ra$, namely for $\alpha> \alpha_m$, there are no connected surfaces homologous to $\ra$, which automatically  forces the holographic entanglement entropy to plateau in the sense described earlier.  Above, we demonstrated this result by explicitly constructing minimal surfaces in the bulk \SAdS{d+1} spacetime; the robustness of  numerical construction rests on explicitly allowing surfaces to relax to minimal ones using a mean curvature flow as explained in Appendix \ref{s:flows}. However, the absence of connected minimal surfaces is in fact necessitated by a certain relational property of minimal surfaces and causal wedges. The latter is a construct that crucially involves the Lorentzian structure since it has to do with causal relations between points. We now proceed to explain the basic physics behind this connection, leaving the technical details for a separate discussion \cite{Hubeny:2013fk}.

Given a boundary region $\ra \subset \Sigma$, one can construct a causal wedge (which, following \cite{Hubeny:2012wa}, we denote by $\cwedge$) associated with it in the bulk. The construction involves first identifying the boundary domain of dependence $\domd$ of $\ra$, which for the rigid boundary geometry is simply the region where one can reliably Cauchy-evolve initial data laid down on $\ra$. If one has full knowledge of $\rho_\ra$ then one can compute correlation functions of all local operators  inserted in $\domd$; this is just a basic statement of causality in relativistic quantum field theories. We then construct the causal wedge $\cwedge$ in the bulk, defined  as the set of points in ${\cal M}$ which can communicate with and simultaneously be communicated from  $\domd$; in other words, through which there exists a causal curve which starts and ends in $\domd$. The reader interested in the precise definitions is invited to consult \cite{Hubeny:2012wa}. 

Since the causal wedge is constructed purely based on where causal curves reach into the bulk, one might naively imagine that for a simply-connected boundary region $\ra$ the causal wedge has trivial topology. This turns out not to be the case \cite{Hubeny:2013fk}. While  the causal wedge for a simply connected region has to be simply connected (a consequence of topological censorship), it can have `holes'. This implies that $\cwedge$ possesses non-contractible $d-2$ spheres on its boundary. Intuitively, this can happen because of geodesic trapping, as we now explain. The boundary of the causal wedge is generated by null geodesics. Bulk  spacetimes which possess  null circular orbits (arising from a sufficiently strong gravitational potential) cut off some of these generators, while allowing some others to fly-by and thereby create a hole in the wedge. Intuitively, the play-off is always between gravitational attraction and centrifugal repulsion and for a wide family of geometries one is able to engineer these so as to obtain the desired effect. Happily for us, the simplest class of examples where such a feature can be exhibited explicitly are the \SAdS{d+1} geometries, wherein for large enough region $\ra$ (which we denote by $\alpha>\alphaCW$),  the causal wedge $\cwedge$ exhibits a hole \cite{Hubeny:2013fk}.

The presence of non-trivial topology has an important implication for the minimal surfaces $\mins{}$.  This is at first sight surprising, as one might have naively thought that since minimal surfaces are defined in the Riemannian section (constant-time slice), they ought to be utterly ignorant of any causal argument which is intrinsically Lorentzian. This naive reasoning is however too quick: One can show that all extremal surfaces $\extr{\ra}$ (of which the minimal surfaces $\mins{\ra}$ considered herein are a special case) must lie outside (or at best on the boundary of) the causal wedge \cite{Hubeny:2012wa,Wall:2012uf,Hubeny:2013fk}. 
Here by `outside' we mean `not within' -- in other words an extremal surface $\extr{\ra}$ must penetrate deeper into the bulk than the corresponding causal wedge $\cwedge$. The result can be intuitively understood by realizing that a minimal surface wants to minimize its area and thus wants to spend much of its existence deep in the interior where it can minimize the red-shift factor of AdS (which augments the area element of the directions along the boundary).  This effect is less dominant for lower-dimensional surfaces, and null geodesics can additionally  offset the large spatial distances by large temporal ones.\footnote{
Note that in comparison to pure AdS geometry, both null geodesics and extremal surfaces are nevertheless pushed towards the boundary by the gravitational potential well of a deformed bulk spacetime.  This in effect  means that both entanglement entropy and causal holographic information defined in \cite{Hubeny:2012wa} grow with positive mass deformations of the spacetime.}
Hence generically the causal wedge boundary is nestled between the extremal surface and the AdS boundary.
As an aside, this statement forms a crucial ingredient in the arguments of \cite{Czech:2012bh} who argue that the holographic dual of the reduced density matrix $\rho_\ra$ must comprise of a bulk region that is larger than the casual wedge $\cwedge$.\footnote{
{\em Caveat:} The statements made above should be viewed with suitable caution, as they require some work to be established rigorously.  For further discussion and a proof of the nesting property of causal wedges and extremal surfaces (modulo some technical assumptions), we refer the reader to \cite{Hubeny:2013fk} (see also \cite{Wall:2012uf} for a related discussion in causally trivial spacetimes).}

We are now in a position to explain the behaviour found in \S\ref{s:shd}. For $\alpha > \alphaCW$ the ability of the causal wedge to develop holes implies that we should anticipate a corresponding change in the minimal surface. Indeed $\mins{\ra}$ cannot pass into the hole in $\cwedge$ while remaining connected to the boundary, because of the nesting property: in order to do so, it would have to pass through the causal wedge. This means that it either lies completely outside the causal wedge or is contained entirely within the hole! One fact which is somewhat obvious by causality is that the hole in the causal wedge must lie outside the black hole event horizon. For static black hole spacetimes we have argued that the bifurcation surface is a candidate extremal surface in \S\ref{s:s2d}. So in the presence of the hole in the wedge we can have a minimal surface component on the far side of the black hole (hence outside $\cwedge$) and another component being simply the bifurcation surface. In fact, as in the BTZ discussion, the homology constraint imposed upon the RT prescription by the Araki-Lieb inequality necessitates both components. Hence the only surfaces that are both minimal and homologous to $\ra$ are the disconnected surfaces $\mins{2}$ described in \S\ref{s:shd}. 

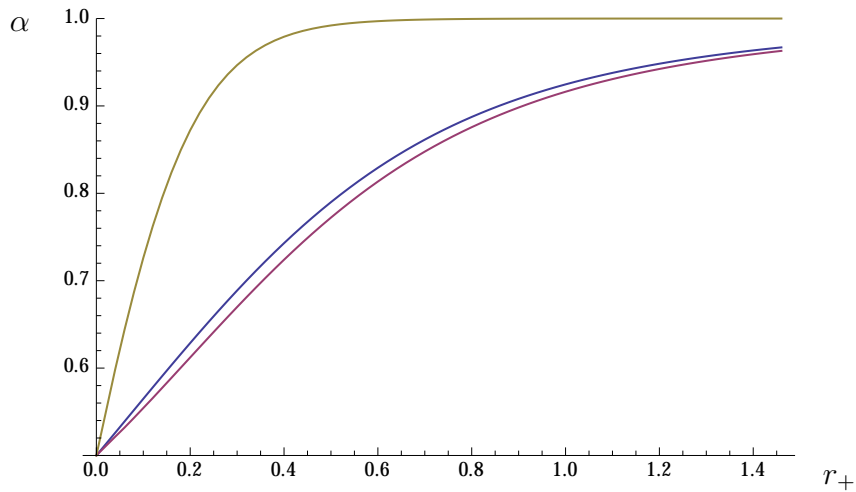
\begin{figure}[h!]
\begin{center}
\includegraphics[width=3.8in]{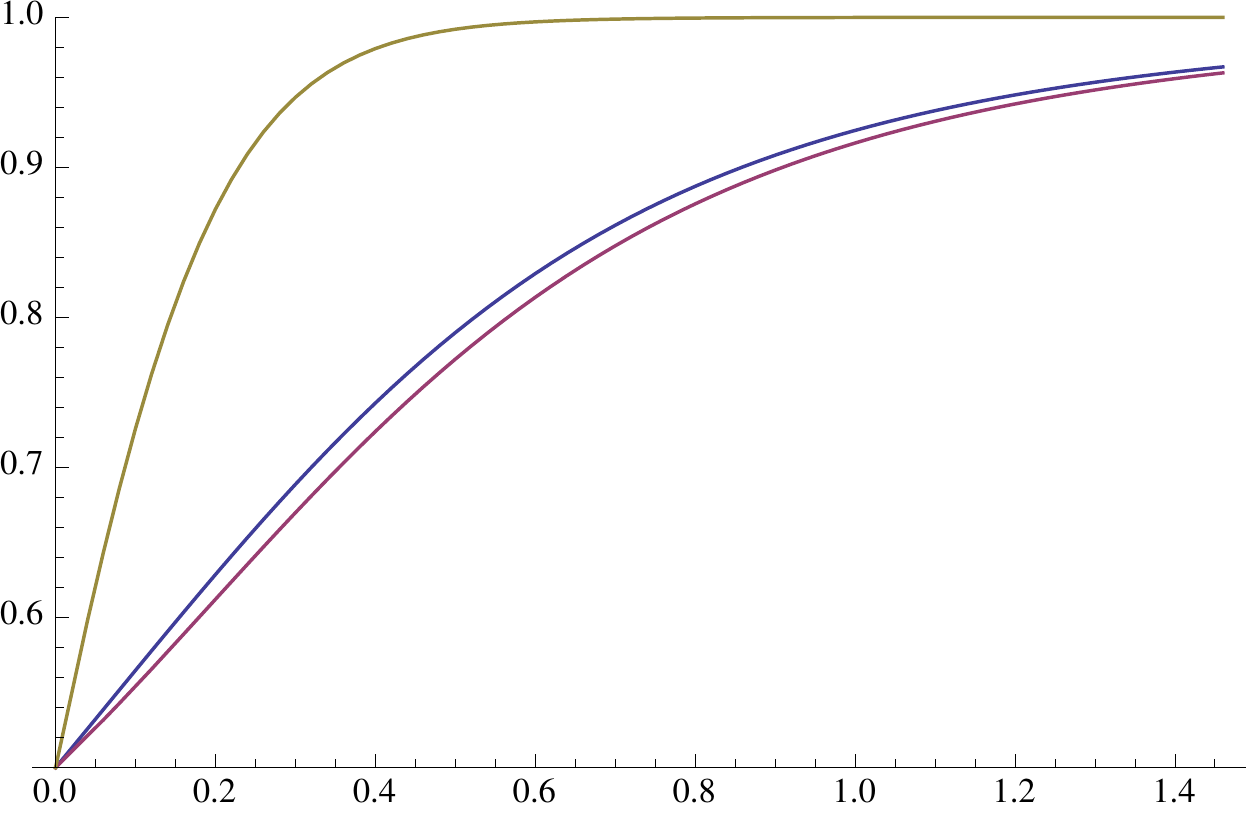} 
\begin{picture}(0,0)
\setlength{\unitlength}{1cm}
\put (-10.5,6){$\alpha$}
\put(0.2,0){$\rh$}
\end{picture}
\caption{
The critical values of $\alpha$ encountered in our discussion as a function of $\rh$.
We plot the largest fraction of the boundary region characterized by $\alpha$ for which 
(i) a connected minimal surface exists $\alpha_m$ (blue), (ii) for which disconnected/connected exchange dominance $\alpha_\trans$ (purple) and, (iii) the critical $\alpha$ for which the causal wedge develops holes $\alphaCW$ (olive) in the \SAdS{5} spacetime.
}
\label{f:alphacritcal}
\end{center}
\end{figure}
We demonstrate explicitly that $\alpha_\trans \le \alpha_m \le \alphaCW$ as one anticipates for first order transitions for \SAdS{5} in \fig{f:alphacritcal}.
As the figure indicates, the causal wedge pinch off happens at $\alpha = \alphaCW$ which is significantly larger than the bound  $\alpha_m$ on existence of corresponding extremal surfaces.  In other words, the connected minimal surface ceases to exist long before this is necessitated by the topology of the causal wedge.  While this may seem to somewhat weaken our arguments regarding the utility of the causal wedge, we should note that the pinch-off value $\alpha = \alphaCW$ is a rather weak bound from the causal wedge standpoint as well: already before the causal wedge pinches off, its geometry precludes the requisite minimal surface.  This is because prior to the pinch-off the opening becomes too long and narrow to admit any extremal surface.  To support the latter, the neck would have to accommodate a sufficient `flare-out' shape, as explained in \sec{s:entplatd}, which  occurs for $\phA$ smaller by $\sim \CO(\rho_0 - \rho_+)$, where $\rho_0$ is the $\rho$-radius of null circular orbit and $\rho_+$ gives the horizon size.  Hence a more sophisticated analysis which takes into account the shape of the causal wedge apart from its topology would provide a much stronger bound on $\alpha_m$.  It would interesting to examine in detail precisely how close we can get to $\alpha_m$, but we leave this for future exploration.

\section{Minimal versus extremal surfaces and the homology constraint}
\label{s:homol}

So far, we have examined the properties of the entanglement entropy as given by the RT proposal involving minimal surfaces (which is a well-defined problem if we can look for our surfaces in a Riemannian geometry).  This allowed us to argue in \sec{s:general} that the entanglement entropy for a certain region $\ra$, measured in a fixed total density matrix $\rho_\Sigma$, is continuous as a function of parameters $\alpha$ describing the region $\ra$.  
On the other hand, the general prescription for holographic entanglement entropy must be formulated in a fully covariant fashion.  In other words, in absence of a geometrically-preferred bulk spacelike slice (which for static cases we could take to be a constant-$t$ slice where $\frac{\partial}{\partial t}$ is the timelike Killing field), we need to have a prescription which does not rely on a specific coordinate choice.
This motivated HRT  \cite{Hubeny:2007xt} to propose the extremal surface prescription \req{HRTprop}, which is fully covariant and well-defined for arbitrary time-dependent asymptotically-AdS bulk geometries.  It has henceforth been assumed that the two prescriptions, RT and HRT, are equivalent for static geometries.  In this section we wish to revisit this assumption, in conjunction with  the closely-related issue of the homology requirement.

Let us start by specifying the prescriptions more explicitly, for completeness also adding a recent `maximin' reformulation of HRT by Wall \cite{Wall:2012uf}.
  In each formulation, the surface is required to be anchored on $\entsurf$, satisfy the homology constraint, and in case of multiple surfaces satisfying these criteria, be the minimal-area one.  The differences in the prescriptions enter at the level of constructing the requisite surfaces, and can be summarized as follows:
\begin{itemize}
\item RT \cite{Ryu:2006bv,Ryu:2006ef}: minimal surface on constant-$t$ spacelike slice $\Sigma_t^{\rm bulk}$
\item HRT  \cite{Hubeny:2007xt}:  extremal surface in the bulk $\bulk$
\item Wall \cite{Wall:2012uf}:  minimal surface on any achronal bulk slice ${\tilde \Sigma}$, maximized over all possible ${\tilde \Sigma}$.
\end{itemize} 
While RT is restricted to static spacetimes, Wall's prescription is formulated in causally trivial bulk geometries.  In this context, \cite{Wall:2012uf} proves equivalence between the maximin and HRT prescriptions assuming the null energy condition.  Although the maximin construction is useful for some purposes (for example, it allows \cite{Wall:2012uf} to argue the existence of such surfaces and prove strong subadditivity in the time-dependent context), it is conceptually more complicated since the requisite surface is obtained by a two-step procedure of first minimizing the area on some achronal slice ${\tilde \Sigma}$, and then maximizing the area with respect to varying ${\tilde \Sigma}$.
Moreover, here we wish to consider causally non-trivial spacetimes, so we will henceforth restrict attention to the RT and HRT proposals.

In a globally static geometry, RT and HRT proposals are indeed equivalent, since any extremal surface anchored at constant $t$ on the boundary must coincide with the minimal surface on $\Sigma_t^{\rm bulk}$, cf.\ \cite{Hubeny:2007xt}.\footnote{
It is worth commenting that the proof presented in \cite{Hubeny:2007xt} shows that the HRT proposal reduces to the minimal surface RT proposal in static geometries. The question of whether there are extremal surfaces  not  localized on a constant time slice and having smaller area than the minimal surface is, as far as we are aware, open. One might argue based on Euclidean continuation of static geometries that such surfaces do not play any role in determining $\SA$; we thank Tadashi Takayanagi for useful discussions on this issue. }
However, in a static but not globally static geometry (i.e.\ when there is a global Killing field which is timelike near the AdS boundary but does not necessarily remain timelike everywhere in the bulk), the situation can appear more subtle.

\begin{figure}
\begin{center}
\includegraphics[width=1.5in]{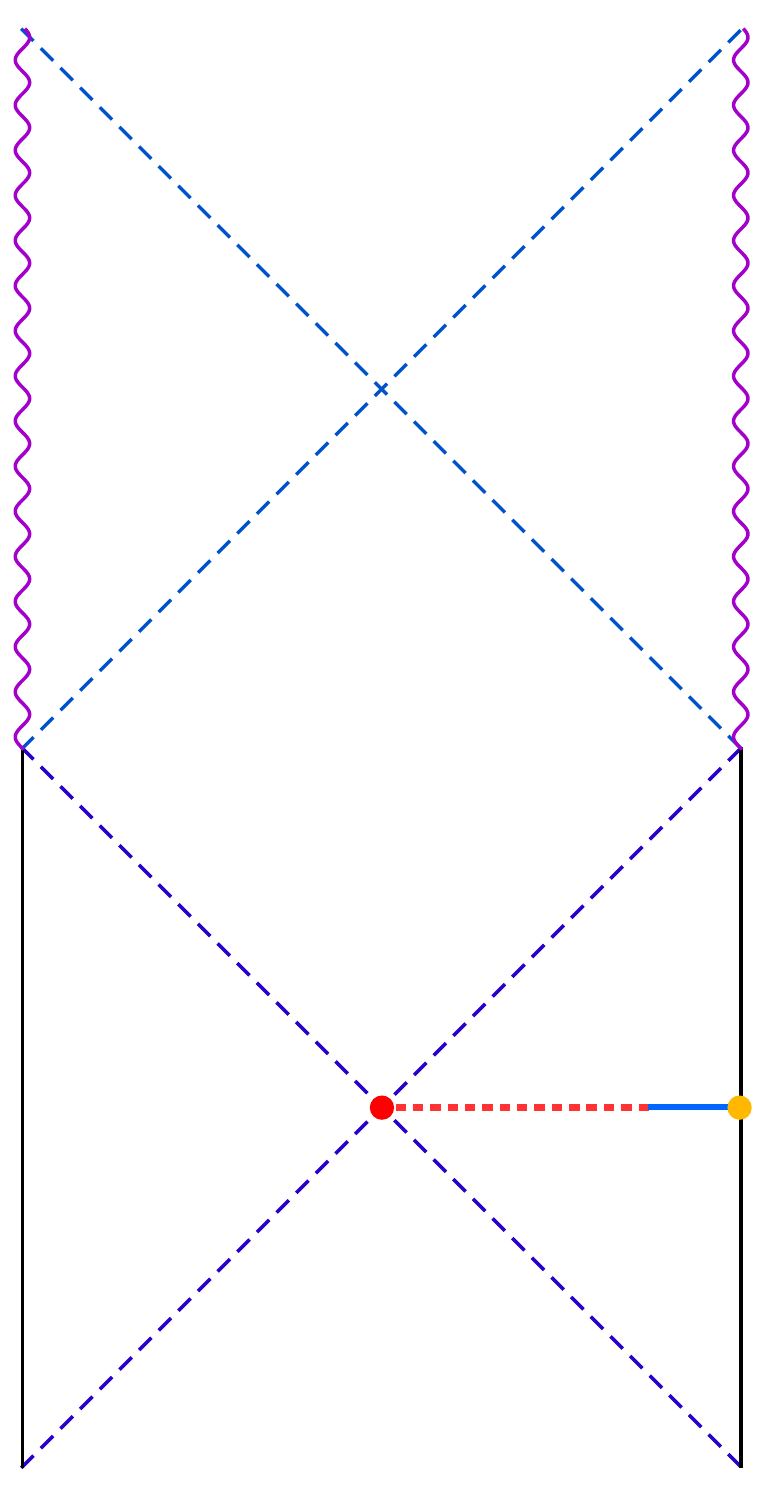} 
\hspace{2.2cm}
\includegraphics[width=1.5in]{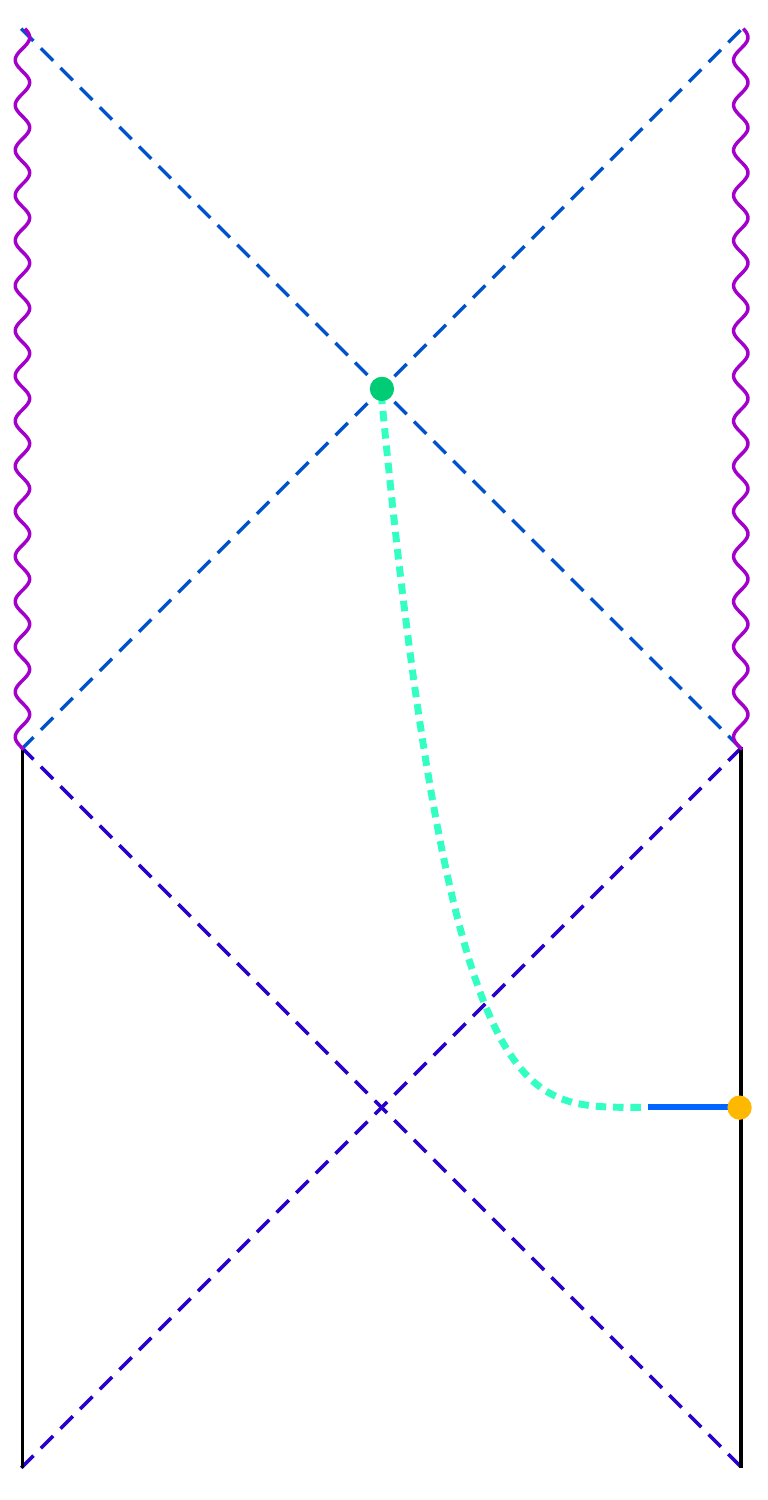} 
\begin{picture}(0,0)
\setlength{\unitlength}{1cm}
\put (-8.4,0.1) {$\vdots$}
\put (-2.1,0.1) {$\vdots$}
\put (-8.4,7) {$\vdots$}
\put (-2.1,7) {$\vdots$}
\put (-11.4,5.5) {$r=0$}
\put (-11.5,2) {$r=\infty$}
\put (-9.3,2.3) {$r=r_+$}
\put (-9.3,4.95) {$r=r_-$}
\end{picture}
\caption{
A sketch of the Penrose diagram for non-extremal Reissner-Nordstrom-AdS black hole (the figure repeats in the vertical direction, indicated by the ellipsis).  The AdS boundaries are indicated by the black vertical lines, the curvature singularities by purple vertical wiggly curves, the horizons by blue diagonal dashed line, the projection of the boundary region $\ra$ by the orange dot, and the projection of the extremal surface by the blue horizontal line.  The two panels distinguish the RT (left) and naive HRT (right) prescriptions in case of disconnected surfaces: in the former the disconnected surface lies at the event horizon (red dot) with the homology region ${\cal R}$ indicated by the red dotted line.  In the HRT case, the disconnected surface is at the Cauchy horizon (green dot), and corresponding ${\cal R}$ then interpolates between this surface and its other boundaries as indicated by the green dotted curve.
}
\label{f:RNAdShomol}
\end{center}
\end{figure}
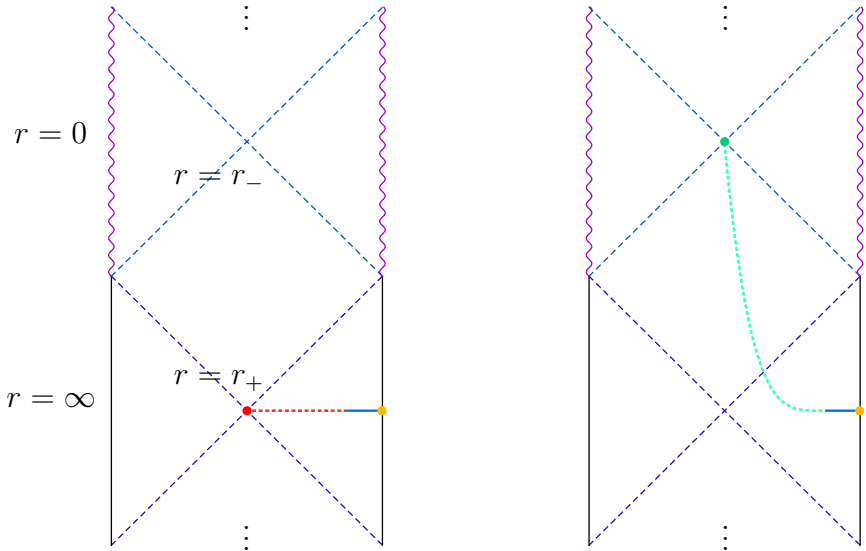

To illustrate the point, let us consider the eternal non-extremal Reissner-Nordstrom-AdS geometry, corresponding to a static charged black hole.  
It has a metric of the form \req{SAdSdmet}, but with $f(r)$ having two positive real roots at $r=r_\pm$ with $r_+$ corresponding to the outer (or event) horizon and $r_-$  the inner (or Cauchy) horizon  with $0<r_-<r_+$.\footnote{
For example, in 4+1 bulk dimensions, 
$f(r) = r^2 + 1 - \frac{M}{r^2} + \frac{Q^2}{r^4} 
= \left(1- \frac{r_+^2}{r^2} \right)  \left(1- \frac{r_-^2}{r^2} \right) \,  (  r^2 + 1 +r_+^2 + r_-^2)$.
}
The causal structure is indicated in the Penrose diagram in \fig{f:RNAdShomol}.  Apart from the bifurcation surface of the event horizon at $r=r_+$ which has zero extrinsic curvature (and therefore is a compact  extremal surface), there is an analogous compact extremal surface at the inner horizon $r=r_-$.  At both surfaces the $\frac{\partial}{\partial t}$  Killing field vanishes (and its norm decreases in spacelike directions and increases in timelike directions), but whereas the area of the surface is minimized to spatial deformations and maximized to temporal ones in the case of event horizon,  it is the other way around for the inner horizon.  

Outside the black hole, the geometry is qualitatively similar to that of \SAdS{}, in the sense that e.g.\ causal wedges for large enough boundary regions $\ra$ will have holes.  This will in turn preclude the existence of connected extremal surfaces anchored on $\entsurf$ and homologous to $\ra$ in this regime, by the argument of \sec{s:causal}.
In such a regime, the RT prescription then instructs us to take the surface homologous to $\rac$, along with the compact surface wrapping the event horizon.  On the other hand, in the HRT prescription, there would a-priori be nothing to prevent us from taking the latter component to wrap the inner horizon instead:  its area is manifestly smaller than that of the event horizon, and it still satisfies the (naive) homology constraint, as indicated pictorially by the green curve in right panel of \fig{f:RNAdShomol}.  In particular, there exists a smooth co-dimension one surface whose only boundaries are the inner horizon, the extremal surface homologous to $\rac$, and the region $\ra$ on the boundary.

Note that if this were indeed the correct prescription, then we would find that, instead of saturating the Araki-Lieb inequality as in \req{ALsaturd}, we would only satisfy it: $\dS$ would still plateau as a function of $\phA$, but at a value which is lower\footnote{
The special case of extremal RN-AdS black hole with $r_-=r_+$ is somewhat more subtle, since there is no bifurcation surface (instead the spacial geometry exhibits an infinite throat).  We think this is a feature rather than a bug, indicative of being at strictly zero temperature; however the HRT prescription can be applied in the same limiting fashion as is commonly done with e.g.\ the Wald entropy of extremal black holes.
}
than the expected value $ \Srho$:
\begin{equation}
\dS (\phA\ge\phA^\trans)
= \frac{\Omega_{d-1} \, r_-^{d-1}}{4\, G_N^{(d+1)}} 
= \left( \frac{r_-}{r_+}\right)^{\! \! d-1} \, \Srho \,.
\label{}
\end{equation}	
Although this result is consistent with the Araki-Lieb inequality,
it is nevertheless at odds with the CFT expectation: for nearly-neutral black holes where $r_- \ll r_+$ we should be close to the thermal value rather than parametrically separated from it!

This observation suggests that we need some modification to the homology constraint specification in \req{HRTprop}: the mere presence of {\it some} smooth surface ${\cal R}$ with the requisite boundaries does not seem to suffice.  One natural restriction, which indeed has already been employed in \cite{Wall:2012uf},
 is to require that ${\cal R}$ be everywhere spacelike.\footnote{
 Alternately, we could require that no component of the extremal surface is allowed to lie in the causal future of ${\cal A}$.
We thank Matt Headrick 
and Don Marolf 
for useful discussions on this point.}
With this additional restriction, the above example would be safely invalidated, since ${\cal R}$ cannot reach from the Cauchy horizon bifurcation surface $r=r_-$ to the AdS boundary $r=\infty$ while remaining spacelike everywhere, as evident from \fig{f:RNAdShomol}.  The only other compact extremal surface which is spacelike-separated from the boundary region $\ra$ and the extremal surface homologous to $\rac$ is the event horizon bifurcation surface, which recovers the thermal answer \req{ALsaturd}, consistently with our expectations.

\begin{figure}
\begin{center}
\includegraphics[width=2.5in]{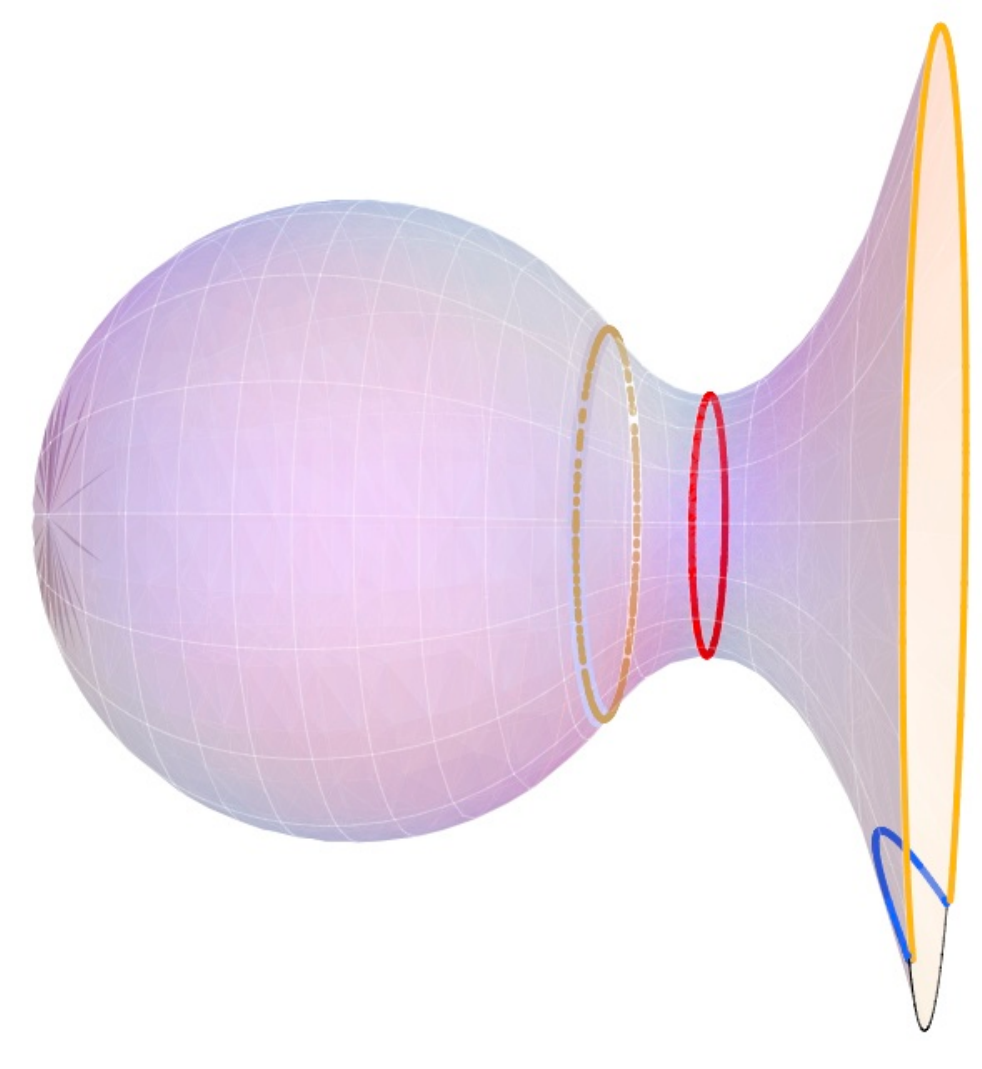} 
\hspace{1.5cm}
\includegraphics[width=3in]{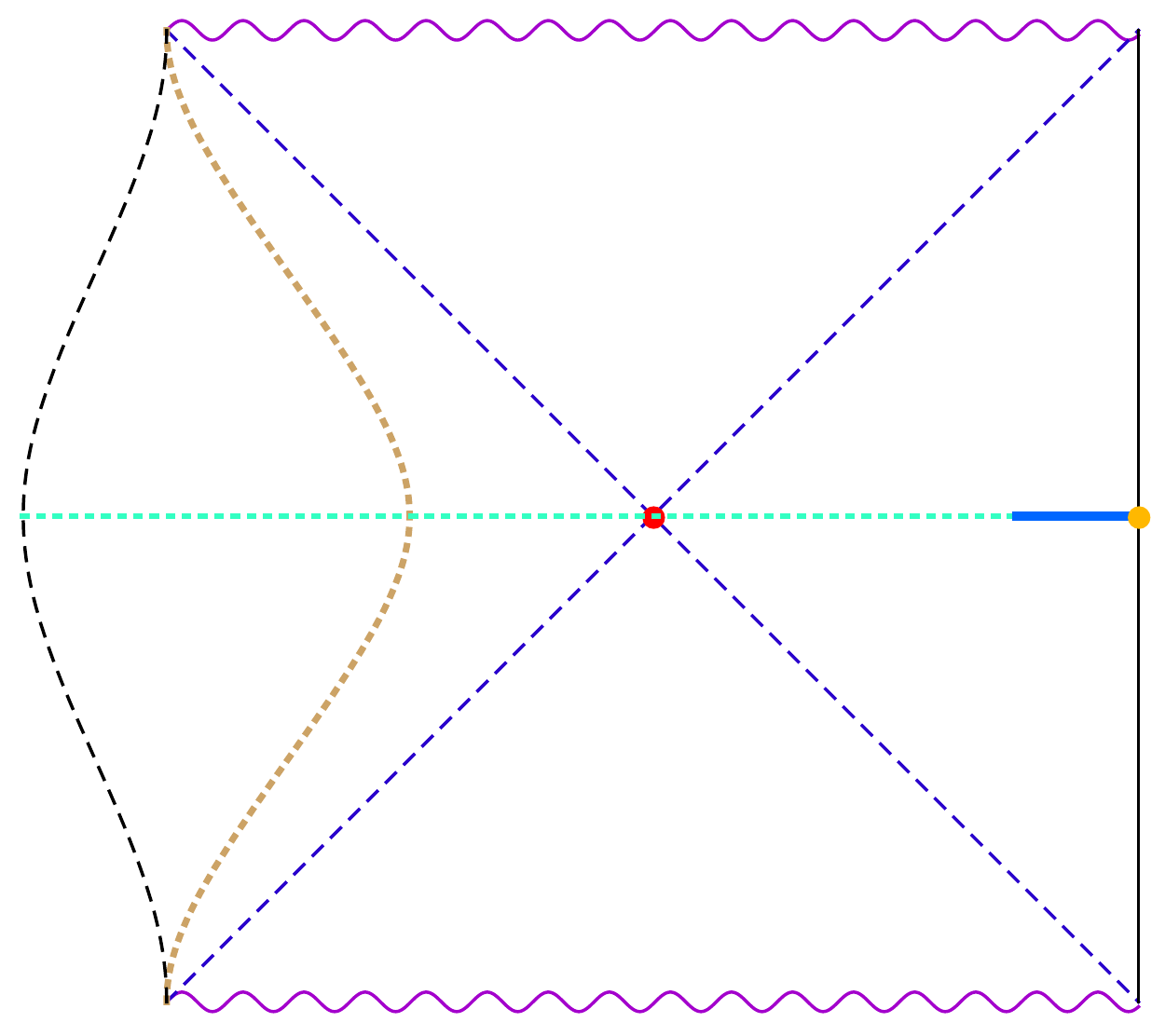} 
\caption{
Sketch of the \SAdS{} `bag of gold' embedding diagram (left) and Penrose diagram (right).  Here the right asymptotic boundary of eternal \SAdS{} geometry is cut-off by a static shell (brown dotted curve), beyond which the spacetime  caps off through a smooth origin as indicated.  The embedding diagram depicts  event horizon bifurcation surface (red curve) and the boundary region $\ra$ (orange curve), at the endpoints of which is anchored the connected part of the extremal surface (blue curve).
The Penrose diagram has the same conventions as in \fig{f:RNAdShomol}, but now the homology surface ${\cal R}$ (dotted green line) goes all the way around the tip.
}
\label{f:SAdSbg}
\end{center}
\end{figure}

However, while the spacelike restriction on the homology constraint recovers the thermal answer for the global eternal charged black hole, there are other geometries where this does not suffice.  As our second exhibit, consider a \SAdS{} `bag of gold' geometry, discussed in e.g.\  \cite{Freivogel:2005qh,Marolf:2008tx}.  This has causal structure and a spatial embedding geometry as sketched in \fig{f:SAdSbg}.  The right asymptotic region, as well as interior of the black hole and white hole, are the same as in the eternal \SAdS{} geometry, but the left asymptotic region is modified by a presence of a shell whose interior has a smooth origin.  Moreover, one can fine-tune  the shell's trajectory such that it remains static -- so the entire spacetime admits a Killing field $\frac{\partial}{\partial t}$.  In the CFT dual, such static geometry describes some equilibrium mixed (though not precisely thermal) density matrix.
   
Let us once again examine the entanglement entropy of a sufficiently large region $\ra$, for which the causal wedge has a hole, so that no connected extremal surface anchored on $\entsurf$ can pass on $\ra$'s side of the black hole.  In the global eternal \SAdS{} spacetime, the homology constraint would then have forced us to take the connected surface on $\rac$'s side of the black hole, along with the bifurcation surface on the event horizon.  In the present case however, while the bifurcation surface is still an extremal surface (cf.\ the red curve in \fig{f:SAdSbg}), its inclusion in the entanglement entropy computation is no longer required by the homology constraint (even including the spacelike restriction):  there exists a smooth spacelike co-dimension 1 region ${\cal R}$ (indicated by the dotted green line in the Penrose diagram of \fig{f:SAdSbg}), wrapping the bag of gold at the given instant in time, whose only boundaries are $\ra$ and the extremal surface for $\rac$ (denoted by the blue curve in \fig{f:SAdSbg}).  In other words, in this geometry,
\begin{equation}
\SA = \SAc \thuss \dS = 0  \qquad \forall \ \phA \,.
\label{ALbagofgold}
\end{equation}	
We stress that while the CFT is not in the precisely thermal state, it is certainly not in a pure state either \cite{Freivogel:2005qh}, so a relation of the form \req{ALbagofgold} is wholly unexpected.\footnote{
However, as indicated in \sec{s:discuss}, similar effect takes place in time-dependent situations involving black hole collapse.}

\begin{figure}
\begin{center}
\includegraphics[width=1.5in]{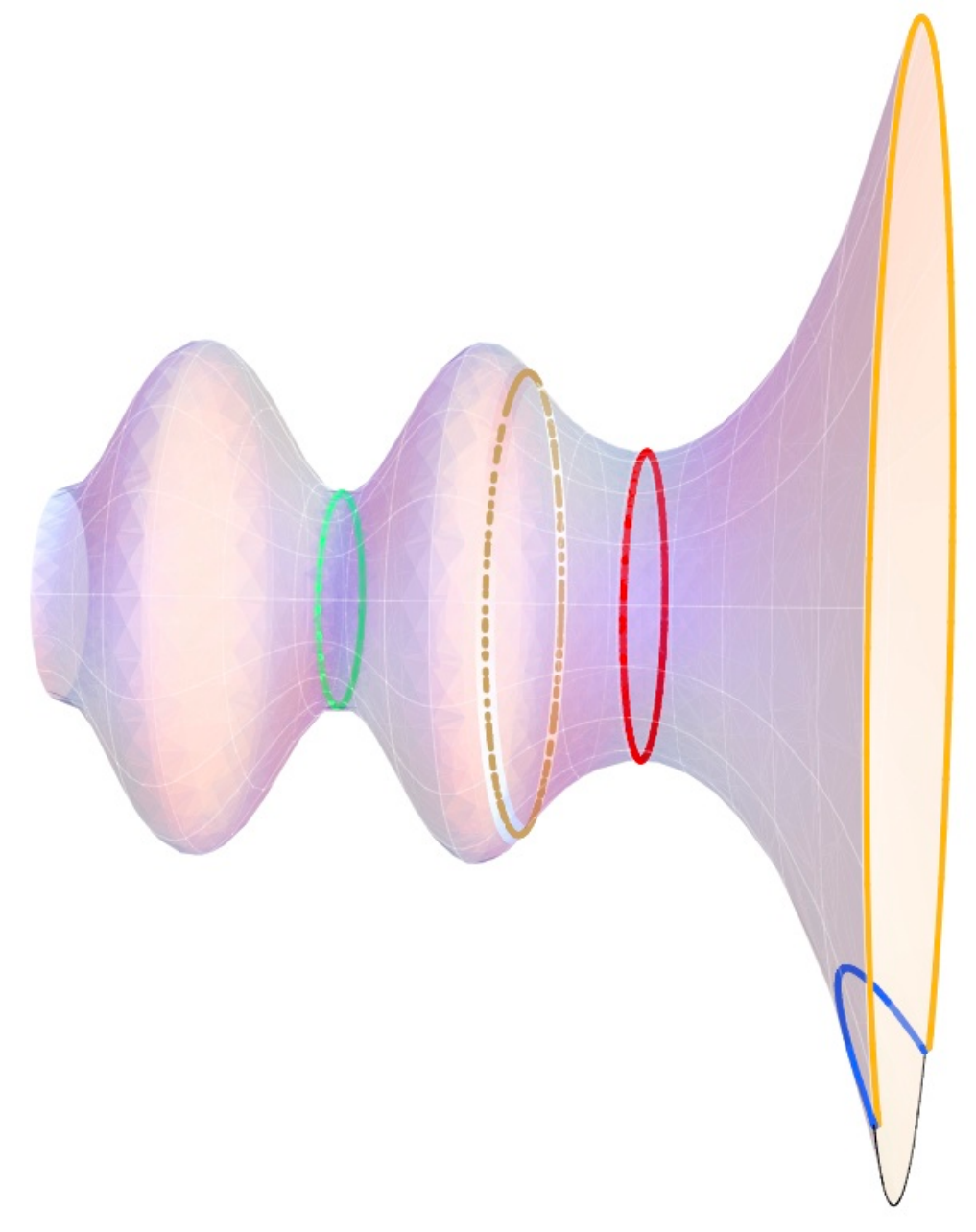} 
\hspace{1cm}
\includegraphics[width=4.7in]{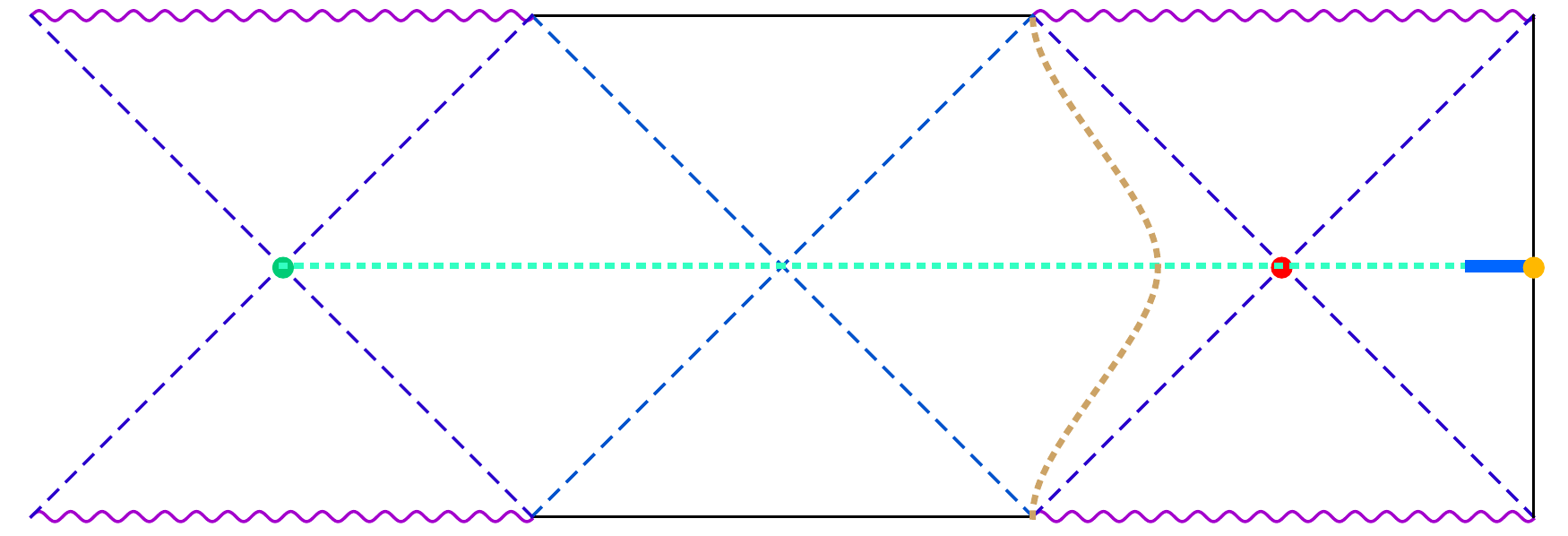}
\begin{picture}(0,0)
\setlength{\unitlength}{1cm}
\put (-12,2) {$\cdots$}
\put (-17.6,2.4) {$\cdots$}
\end{picture}
\caption{
Sketch of the \SAdS{} joined to Schwarzschild-dS geometry across a shell: embedding diagram (left) and Penrose diagram (right), with same conventions as in \fig{f:SAdSbg}.  In contrast to the capped off geometry, however, here we do have a smaller (but non-zero) area extremal surface (green circle/dot) which is required by the homology constraint.
}
\label{f:SAdSSdS}
\end{center}
\end{figure}
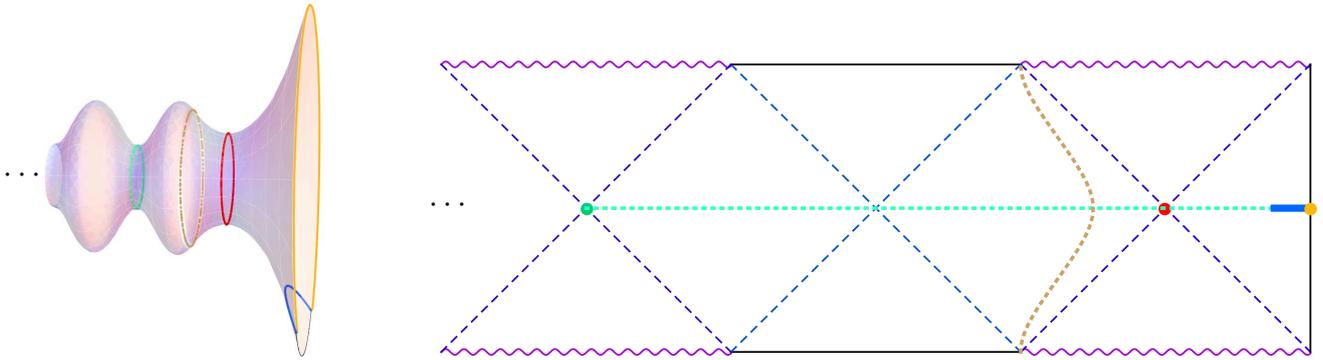

In comparing the RT and HRT prescriptions for the  \SAdS{} bag of gold case, we encounter a slight ambiguity in the RT prescription, namely in what is really meant by a `constant $t$ slice':  strictly-speaking, the static coordinate patch ends at the horizon (which also follows from thinking about the Euclidean section), so the RT prescription should not see any difference between this and the eternal \SAdS{} geometry -- hence predicting that $\dS = \Srho$ for large regions.  On the other hand, if one were instructed to take a full (geodesically complete) time-symmetric slice through the global geometry, one would reproduce the HRT result \req{ALbagofgold}.  We also note that one could consider intermediate cases, with any $\dS \le \Srho$, by putting a Schwarzschild-dS geometry with smaller event horizon ${\tilde r}_+ < \rh$ on the left of the shell, as indicated in \fig{f:SAdSSdS}.  The corresponding embedding diagram for a time-reflection-symmetric slice would not cap off, necessitating the inclusion of a bifurcation surface for this Schwarzschild-dS geometry in the entanglement entropy computation, in order to satisfy the homology constraint.   

We have presented several examples of static (though not globally static) asymptotically AdS geometries for which the RT and HRT proposals might differ, depending on the precise formulation of the homology constraint and of RT's constant time slice.  We have not offered any definitive resolution; this is an interesting and important area which we leave for future exploration.
The specific simple examples we have presented above all have some potentially dubious features: in case of Reissner-Nordstrom-AdS, the Cauchy horizon is unstable, while in the other two cases, the shell is unstable.  Moreover, in all cases, the difference between the prescriptions occurs due to a part of the geometry which is beyond the horizon and thus causally inaccessible to an asymptotic observer.
While this feature might therefore seem rather unappealing, we stress that in general we would be forced into such situation in any case, as long as the entanglement entropy is related to some locally-defined geometric construct, due to the teleological nature of the event horizon.  The examples mentioned above merely illustrate the issues we have yet to confront to fully understand the holographic entanglement entropy prescription.

\section{Discussion}
\label{s:discuss}

We have focused on exploring the behaviour of entanglement entropy $\SA$ under smooth deformations of the entangling region $\ra$ in finite systems. Of particular interest to us is the distinction between the behaviour of holographic field theories, which in the large $c$ (planar limit) and strong coupling limit can be mapped onto the dynamics of classical gravity, versus field theories away from the planar limit. Sharp features in observables are possible in the latter since the planar limit allows one to enter a `thermodynamic regime', as evidenced for example by the thermal phase transitions in finite volume \cite{Witten:1998zw, Sundborg:1999ue, Aharony:2003sx}. 

We focused in particular on the Araki-Lieb inequality which gives a useful measure of the relative entanglement of a region $\ra$ and its complement $\rac$. While this would vanish if the entire system were in a pure state, it carries non-trivial information about the total density matrix in general. We indeed encounter an interesting phenomenon of entanglement plateaux: the Araki-Lieb inequality is saturated for finite system sizes, owing to some non-trivial features of minimal surfaces which compute the holographic entanglement entropy. 

Focusing on thermal density matrices in CFTs we find that in $1+1$ dimensional field theory the Araki-Lieb inequality forces us to modify the expression of entanglement entropy for a large enough region (a point previously noted  in \cite{Azeyanagi:2007bj,Blanco:2013joa}) and provide an  analytic prediction of when the plateau is attained. We also contrast this behaviour of large $c$ theories against low central charge theories. There being very few exact results on the entanglement entropy of thermal CFTs in finite volume, we focused on the available expressions for Dirac fermion (with and without chemical potential). In the $c=1$ case we note the absence of any plateau and the Araki-Lieb inequality is only saturated when the region under consideration (or its complement) is maximal. In higher dimensional holographic examples this no longer is the case, $\dS$ is  forced by virtue of the features of the holographic construction to plateau (see below).

For the main part of this work, we have considered the RT prescription for calculating the holographic entanglement entropy, which is valid for static equilibrium situations.  In this context, one is instructed to work at a given instant in time, and the entanglement entropy computation then involves finding the area of a requisite {\it minimal} surface in the corresponding Riemannian geometry.  Though innocuously simple-sounding, there is a rich set of features associated with what precisely is meant by `requisite'.  For a specified boundary region $\ra$, the boundary of the relevant surface $\mins{\ra}$ must coincide with the entangling surface $\entsurf$, it must be homologous to $\ra$, and in case of multiple such surfaces, it must be the one with smallest area.

It has already been observed previously that the last restriction can cause the entanglement entropy to have a kink (i.e., its first derivative with respect to the parameter $\alpha$ characterizing the region $\ra$ can be discontinuous).  This is because there can exist multiple families of minimal surfaces which can exchange dominance.  Here we have explored this multiplicity further, and discovered that even in the most simple case of global eternal \SAdS{d+1} geometry with $d\ge 3$, in a certain regime of $\alpha$ there is actually an infinite tower of minimal surfaces anchored on the same entangling surface (though only the lowest two have a regime of dominance), seemingly approaching self-similar behaviour.  This curious feature can arise thanks to the dimensionality of the surfaces and compactness of the horizon.  On the other hand, in other regimes of $\alpha$ (namely for sufficiently large region $\ra$), there is only a single, disconnected minimal surface satisfying the homology requirement -- unlike in the $2+1$ dimensional case, a connected minimal surface homologous to $\ra$ simply does not exist.  This novel feature may be understood as a consequence of certain properties of causal wedges discussed in  \cite{Hubeny:2013fk}. 

We note in passing that the distinction between the \AdS{3} and higher dimensional examples is quite reminiscent of the Hagedorn behaviour encountered in the dual field theories. In  \AdS{3} the BTZ black hole saddle point exists for all values of temperature (as does the thermal AdS one), but it becomes sub-dominant at low temperatures. In higher dimensions, the \SAdS{} saddles only exist above a minimum temperature $T_{min}$; they however take over from the thermal AdS saddle at a slightly higher temperature $T_H > T_{min}$ (when the horizon size is comparable to \AdS{} length scale). So at low enough temperatures we are always forced into the confining state. In the context of entanglement entropy the analogous observation is about the existence of connected minimal surfaces in the black hole geometry (we should emphasize that we are always in the deconfined phase to be able to use the black hole saddle). In three dimensions, connected surfaces always exist but fail to be dominant past some critical region size; in higher dimensions they cease to exist past a critical region size given the homology constraint. 

Our discussion was primarily focussed on entanglement entropy and in particular on $\dS$. One could equally have focussed on the mutual information, $I({\cal A},{\cal B}) = \SA + S_{{\cal B}} - S_{{\cal A}\cup {\cal B}}$, for  two disjoint regions ${\cal A}$ and ${\cal B}$. Holographic studies of mutual information also reveal an interesting behaviour: for sufficiently separated regions $I({\cal A}; {\cal B}) =0$. This is due to the fact that the extremal surface $\extr{{\cal A}\cup {\cal B}}$ breaks up into a disconnected surface anchored on ${\cal A}$ and ${\cal B}$ to avoid thin necks. This has been extensively discussed in the literature cf., \cite{Headrick:2010zt} and \cite{Morrison:2012iz, Hartman:2013qma,Shenker:2013pqa} for implications of this construct in the context of probing black hole interiors. For the regions we have considered $\ra$ and $\rac$ are not strictly disjoint, so we have to be a bit more careful; naively the mutual information involves the sum of the entropies of two regions (which we take to be $\ra$ and $\rac$) while the Araki-Lieb inequality constrains their difference, giving $I(\ra;\rac) = 2 \, \SA$ which is UV divergent.\footnote{
We thank the referee for spotting an error in our original statement and for suggesting use of the auxiliary purifying system.} One can however constrain the mutual information between the smaller of $\ra$ or $\rac$ and the auxiliary system $\hat{\rho}_\Sigma$ which purifies the global density matrix $\rho_\Sigma$ (thus $\Srho = S_{\ra\cup\rac} =S_{\hat{\rho}_\Sigma}$) to vanish when Araki-Lieb inequality is saturated. For example one has 
\begin{align}
I(\ra; \hat{\rho}_\Sigma) = \SA + S_{\hat{\rho}_\Sigma} - S_{\ra \cup \hat{\rho}_\Sigma} = \SA + \Srho - \SAc = 
\begin{cases}
0, \quad \quad\;\;\, \text{for}\;\;\alpha <\frac{1}{2}\\
2\, \Srho, \quad \text{for}\;\; \alpha > \frac{1}{2}
\end{cases}
\label{}
\end{align}
In deriving this we used the fact that $S_{\ra \cup \rac \cup \hat{\rho}_\Sigma} = 0$ by definition of the purifying degrees of freedom. In the case of the global thermal density matrix $\hat{\rho}_\Sigma$ is the thermofield double which lives on the second asymptotic region of the eternal black hole Penrose diagram.\footnote{
It is worth mentioning that the saturation of the Araki-Lieb inequality in large $c$ theories implies that the degrees of freedom in ${\cal A}$ admit a canonical split into two uncorrelated parts: one that carries all the entanglement with ${\cal A}^c$ and another that carries the macroscopic entropy of $\rho_\Sigma$, cf., \cite{Zhang:2011ab}.  We thank Matt Headrick for suggesting this interpretation and other comments regarding  Araki-Lieb inequality saturation.}

While these causal wedge considerations already by themselves guarantee that the entanglement entropy $\SA(\alpha)$ cannot be a smooth function for thermal states in higher dimensions, we have argued that, if defined by a minimization procedure as in the RT prescription, the entanglement entropy must nevertheless be continuous.
Our argument crucially used minimality.  In particular, in the HRT prescription of computing entanglement entropy via a smallest area {\it extremal} surface, the continuity argument given in \S\ref{s:general} does not apply.   It would be interesting to  explore whether sufficient time-dependence can provide counter-examples to continuity or whether the one can generalize the proof of continuity to the Lorentzian context; we leave further investigation of this  issue for the future \cite{Hubeny:2013uq}. 

Note that in this collapsed black hole context, the homology requirement is satisfied automatically, though as explained in  \cite{Hubeny:2013uq}, that itself leads to very curious feature of entanglement entropy:  it can distinguish between an eternal black hole and a collapsed one, arbitrarily long after the collapse had taken place \cite{Takayanagi:2010wp}.  In this sense, while in the AdS/CFT context we are used to classical bulk surfaces providing at best only some coarse-grained CFT information, 
 the homology requirement induces a fine-grained aspect to the entanglement entropy observable.

While in the time-dependent setting mentioned above, the RT prescription is meaningless and therefore cannot be compared to the extremal surface generalization of HRT,
we have seen that there are subtle differences between the RT and HRT proposals even is static situations.
Indeed, a similar `fine-grained' quality as mentioned above is manifested in the entanglement entropy for e.g.\ the \SAdS{} bag of gold geometry discussed in \sec{s:homol}:  there the entanglement entropy (as given by the HRT prescription) can easily distinguish the differences in the geometry behind the horizon, on the other side of the Einstein-Rosen bridge.  Other known diagnostics of the CFT density matrix not being precisely thermal typically require rather sophisticated machinery, such as detecting lack of periodicity on the Euclidean time circle of the analytically continued solution.  
We find it remarkable that entanglement entropy does the job so easily.  This observation of course crucially hinges on the precise formulation of the homology constraint, which deserves to be understood better.

\acknowledgments 
It is a pleasure to thank Jan de Boer, Juan Jottar, Hong Liu, Don Marolf, Tadashi Takayanagi and especially Matt Headrick for various illuminating discussions. We would also like to thank Hong Liu, Matthew Headrick and Tadashi Takayanagi for their detailed comments on a draft of this manuscript.  VH and MR would like to thank CERN, ITF, Amsterdam and ICTP for hospitality during the early stages of this project. In addition MR would like to acknowledge the hospitality of Technion, Israel during the ``Relativistic fluid dynamics and the gauge gravity duality'' workshop as well as the organizers of the ``2nd Mediterranean Conference on Classical And Quantum Gravity'' Veli Losinj, Croatia for their hospitality during the concluding stages of this project. HM is supported by a STFC studentship. 
 VH and MR are supported in part by the STFC Consolidated Grant ST/J000426/1.

\appendix
\section{Mean Curvature Flow}
\label{s:flows}

In this appendix we summarize the algorithm we used to construct the minimal surfaces in \SAdS{d+1} for $d \geq 3$. While one can directly numerically solve the Euler-Lagrange equations of \eqref{lagmin} directly, it is especially useful to consider alternate strategies to ensure that we have obtained all the surfaces of interest. 

For conceptual and computational purposes, it is useful to view minimal surfaces as the endpoint of a flow of surfaces of decreasing area. This is analogous to a gradient descent algorithm for finding a local minimum of a function, for which one considers a point moving with velocity equal to minus the gradient of the function. This gives a curve along which the function monotonically decreases, and exponentially decays to the minimum as a function of flow time  $t_{flow} \equiv \ft \to\infty$.

In the analogous process for minimal submanifolds, the function to be minimized becomes the area functional, and the flow velocity vector becomes a vector field (which may be taken to be normal) defined on the submanifold. The appropriate vector field is given by the \emph{mean curvature} of the submanifold, and the evolutionary process is known as mean curvature flow.

We review the required differential geometric technology, and outline an algorithm for computing the process in the case of codimension-one minimal surfaces.\footnote{
The minimal surfaces relevant for computing entanglement entropy using the prescription of \cite{Ryu:2006bv} are co-dimension two in the bulk. In static spacetimes however we can localize on a constant time bulk hypersurface, leading thus to a search of co-dimension one surfaces on the slice.}

\subsection{Some geometry}

Let $\X$ be a smooth $n$-dimensional Riemannian manifold (e.g., $\X$ for situations of interest is a constant time surface in the bulk spacetime ${\cal M}$), with metric $g$ and Levi-Civita connection $\nabla$.  Further, let $\M \subset \X$ be a smooth $m$-dimensional submanifold (with $m<n$), defined by embedding $\psi:\M\to\X$. The  induced metric  on $\M$ is inherited from $\X$ as $\gamma = \psi^\ast g$ given this embedding. Vectors at a point on the submanifold can be decomposed into components tangent and normal to $\M$: $U=U^\top+U^\bot$.

The natural structures associated with  $\M \subset \X$ are in terms of the intrinsic and extrinsic geometry of the embedding, characterized by the induced metric $\gamma$ (the {\em first fundamental form}) and the extrinsic curvature $\sff$ (the {\em second fundamental form}). Formally, given vector fields, say $U$ and $V$, tangent to $\M$, one can decompose the covariant derivative:
\begin{equation}
  \nabla_U V=(\nabla_U V)^\top+(\nabla_U V)^\bot=\mathcal{D}_U V+ \sff(U,V)\,,
\end{equation}
which effectively defines $\mathcal{D}$ and $\sff$. The former is the Levi-Civita connection of the induced metric on $\M$, while the latter is a normal-valued symmetric bi-linear form. It is worth emphasizing that 
the second fundamental form depends only on the vectors at the point, so it is a tensor on $\M$. A particularly important object is obtained by taking its trace $\traceK$, giving a normal vector field on $\M$: this is the \emph{mean curvature field}.\footnote{
This definition of the mean curvature is not entirely universal: in some conventions the sign differs, while in others the trace is divided by a factor of $m$, so that $\traceK$ is genuinely the mean of the eigenvalues of $\sff$ (which incidentally define the principal curvatures of the submanifold).} In components, using coordinates $x^a$ intrinsic to $\M$, and $X^\alpha$ on $\X$, we have $\gamma_{ab} = g_{\alpha \beta}\, \partial_a X^\alpha \, \partial_b X^\beta $, and
\begin{equation*}
h^{\alpha\beta} = \gamma^{ab}\, \partial_a X^\alpha \, \partial_b X^\beta \,, \qquad 
K_{\alpha\beta}^{\;\;\;\;\gamma} = h_\alpha^{\;\;\mu} \, h_\beta^{\;\; \nu} \, \nabla_\nu\, h_\mu^{\;\;\gamma}
\,, \qquad 
\traceK^\alpha = h^{\mu\nu} \, K_{\mu\nu}^{\;\;\;\alpha}.
\label{}
\end{equation*}	
Here $h$ is the projection onto vectors tangent to $\M$. Note that the extrinsic curvature tensor $K_{\alpha\beta}^{\;\;\;\;\gamma}$ is symmetric in its lower indices which lie tangent to $\M$, while the upper index is transverse.

We are now in a position to make precise the geometric interpretation of the mean curvature field alluded to above, that it is minus the gradient of the area functional on the space of submanifolds.

Let us consider a smooth family of embeddings $\Psi:I\times\M\to\X$, where $I$ is some interval parameterized by a flow time $\ft$. For each $\ft \in I$, $\psi_\ft=\Psi(\ft,\cdot):\M\to\X$ is an embedding. Let $A$ be the area of the embedded submanifold:
\begin{equation}
 A(\ft)=\int_\M \omega_\ft,
\end{equation}
where $\omega_\ft$ is the induced volume form on $\M$ at time $\ft$.

Let $V$ be the vector field given by the local velocity of the surface in this deformation, so $V=\Psi_\ast \frac{\partial}{\partial \ft}$. One should think of $\Psi$, or $\psi_\ft$, as defining a curve through the space of embeddings, an infinite-dimensional manifold whose tangent space at $\psi$ is given by the set of $T\X$-valued vector fields on $\psi$. Then $V$ is naturally the tangent vector to the curve.

We now have enough state our key result:
\begin{equation}
 \frac{\dee A}{\dee \ft}=-\int_\M g(\traceK ,V)\, \omega_\ft.
\end{equation}
In the space of embeddings, this is the natural inner product between $V$ and $-\traceK$, which makes precise the analogy with a gradient. Note that, as is to be expected, only the normal component of $V$ contributes,  since the tangent component is `pure gauge', describing how parameterization of the embedding evolves, and not the shape of the submanifold itself.

\begin{proof}
 The method of proof is to compute $\frac{\partial}{\partial \ft}$ of the induced volume form on $\M$, and we will show it is equal to $(-g(\traceK,V)+\mathcal{D}\cdot V^\top)\omega_\ft$. Since the last term is a total derivative, the result will follow, provided that the variation is constrained to a compact region, or we restrict the variation to be normal (so $V^\top=0$). For ease of computation, we work in a chart of normal coordinates $x^i$ on $\M$ at the point $p$, with respect to the induced metric at time $\ft=0$: $\gamma(\partial_i,\partial_j)=\delta_{ij}$, and $\mathcal{D}_{\partial_i}\partial_j=0$. Let $e_i=\Psi_\ast\partial_i$. For later use, we note that since Lie brackets commute with push-forward, $[e_i,V]=0$, so $\nabla_V e_i=\nabla_{e_i}V$.
 
 The induced volume form is given by $\omega_\ft=\sqrt{\det(g(e_i,e_j))}\; \dee x^1\wedge\cdots\wedge\dee x^m$,
 so its derivative at the point $p$ and $\ft=0$ is
\begin{align}
\frac{1}{2}\frac{\partial}{\partial\ft}(g(e_i,e_i)) \; \omega_0,
\label{}
\end{align}
where we have used the fact that the derivative of $\det A$ is $\det A \, \mathrm{tr}(A^{-1}\dot{A})$, and that $\dee x^1\wedge\cdots\wedge\dee x^m=\omega_0$ at $p$ in our normal coordinates. Now, again computing at the point $p$ and $\ft=0$,
\begin{align*}
\frac{1}{2}\, \frac{\partial}{\partial \ft}g(e_i,e_i)  &= \frac{1}{2} V(g(e_i,e_i))\\
      &=  g(\nabla_V e_i,e_i)\\
      &=  g(\nabla_{e_i} V,e_i)\\
      &=  e_i(g(V,e_i))- g(V,\nabla_{e_i}e_i)\\
      &=  \partial_i V^\top_i - g(V,\mathcal{D}_{e_i}e_i+\sff(e_i,e_i))\\
      &= \mathcal{D}\cdot V^\top - g(V,\traceK).
\end{align*}
In the final line we have used the fact that we have normal coordinates, so $\partial_i=\mathcal{D}_i$, and the sum over $i$ reduces to the trace.
\end{proof}

\subsection{The algorithm}

With the necessary mathematical formalism in hand, we now describe an algorithm for implementing the mean curvature flow, specializing for simplicity to the case of codimension-one surfaces. In this case, having chosen a direction for a normal, the mean curvature vector field reduces to a scalar field, since the normal space is one-dimensional.

The method used is based on a level-set approach: the surfaces are described by the sets
\begin{equation}\label{sigmadef}
 \M=\{x\in\X| \;\; \Phi(x)=c \}
\end{equation}
for some function $\Phi$ and some constant $c$, and the evolution of the function $\Phi$ is what we shall model. This has several advantages:
\begin{itemize}
 \item There is no requirement to pick a parameterization of the surface, so no problem with `gauge fixing'.
 \item A whole family of surfaces can be modeled at once, by computing the evolution for $\Phi$ and picking different level sets.
 \item The algorithm elegantly allows for changes in topology: under mean curvature flow, surfaces can split into several parts, and this causes no trouble for the level set method.
\end{itemize}

The definition of the flow gives the normal speed of a point on the surface as it evolves: $n(\frac{\dee x}{\dee \ft})=\traceK$, with $n$ being the normal to $\M$. Provided that $\dee\Phi\neq0$, we also have a simple way of computing a unit normal, $n=\frac{\dee\Phi}{\|\dee\Phi\|}$. To deduce the evolution of $\Phi$, differentiate the definition~\eqref{sigmadef} of a surface $\M$:
\begin{align*}
 0  &=  \frac{\partial\Phi}{\partial \ft}+\dee\Phi\!\left(\tfrac{\dee x}{\dee \ft}\right) =	\frac{\partial\Phi}{\partial \ft}+\|\dee\Phi\|  \,  n\!\left(\tfrac{\dee x}{\dee \ft}\right) =	\frac{\partial\Phi}{\partial \ft}+\|\dee\Phi\| \; \traceK \,.
\end{align*}
The only ingredient which remains is to compute $\traceK$ using only information from $\Phi$. Luckily, this is given by a simple formula in terms of the normal ${\tt k } = -\nabla \cdot n$. To see this,  take $n$ to be the unit one-form, normal to the surface, but also defined, and unit, off it. Extend to an orthonormal basis\footnote{
We denote the `raised index' normal vector by $n^\#$, so that $g(n^\#,U)=n(U)$} $\{e_i,n^\#\}$. Then:
\begin{align*}
 \traceK &=	n(\nabla_{e_i}e_i) = 	e_i(n(e_i))-(\nabla_{e_i}n)(e_i)\\
	&=	-\nabla\cdot n + (\nabla_{n^\#}n)(n^\#)\\
	&=	-\nabla\cdot n + g(n,\nabla_{n^\#}n)\\
	&=	-\nabla\cdot n + \frac{1}{2}n^\#(g(n,n)) =	-\nabla\cdot n.
\end{align*}
The term $e_i(n(e_i))$ in the first line vanishes because $n(e_i)=0$ everywhere, and the last equality follows because $g(n,n)=1$ everywhere.

This gives us our final evolution equation for $\Phi$:
\begin{align}
 \frac{\partial\Phi}{\partial \ft} &= 
 \|\dee\Phi\| \; \nabla\cdot\left(\frac{\dee\Phi}{\|\dee\Phi\|}\right) =	\sqrt{g^{\gamma\delta}\, \partial_\gamma\Phi\, \partial_\delta\Phi}\; \frac{1}{\sqrt{g}}\; \partial_\alpha\left(\frac{\sqrt{g}\, g^{\alpha\beta}\, \partial_\beta\Phi}{\sqrt{g^{\gamma\delta}\, \partial_\gamma\Phi\partial_\delta\Phi}}\right) .
\end{align}
This is a quasilinear parabolic PDE in $\Phi$, which can be solved numerically by standard methods.

For the computations in this paper, only spherically symmetric surfaces were considered, so $\Phi$ was a function of two spatial coordinates $\ph$ and $r$ (or another suitably redefined radial coordinate). The PDE was solved with difference methods, with Neumann boundary conditions at the poles of the sphere ($\ph=0,\pi$) to keep the surfaces regular there, and mixed boundary conditions at radial boundaries set by the desired asymptotics.


\section{Near-horizon behaviour of minimal surfaces}
\label{s:nhorizon}

An analytic solution of the equations of motion in a near-horizon approximation explains the existence of an infinite self-similar family of minimal surfaces, wrapping the horizon multiple times. We find that this is generic near any minimum of the radius of the ${\bf S}^{d-1}$, such as the Einstein-Rosen bridge of a non-extremal black hole, and is characterized entirely by the number of dimensions,\footnote{
Here we will keep to spatial codimension-one surfaces, but the argument carries over essentially unchanged for surfaces of any dimension $m$, wrapping a ${\bf S}^{m-1}$ in an $SO(d)$ symmetric space}, and a single parameter $\xi \equiv \rh\, \kappa$, where $\kappa$ is the surface gravity (which is related to the black hole temperature as $T  = \frac{1}{4\pi}\, \kappa$).

For the analysis we use a modified radial coordinate $z$, regular at the event horizon, defined by $r=r_+ + \frac{1}{2}\, \rh\, \xi \, \,z^2$, and parametrize the surface by $\ph$ throughout. Dots will denote differentiation with respect to $\ph$.

We invoke two approximation schemes. The naive linearization holds away from the poles, but breaks down close the $\ph=0,\pi$ where $\dot{z}$ grows without bound. Near the poles, we use a different approximation, reducing the problem locally to flat Euclidean space. We will find that these two approximation schemes have overlapping domains of validity, so the complete behaviour can be described by matching the asymptotics in each regime. This predicts a tower of arbitrarily many branches of the surface wrapping the horizon, with a known relationship between the distances from the horizon of successive surfaces.

Schematically: we start at the North pole $\ph=0$, and solve in the linear regime towards the South pole $\ph=\pi$. There we change to the near-pole regime, in which there will be a turning point at a minimum angle $\Delta$ away from the pole, at $z=z_0$. These constants are both computed from matching to the linearized solution. We then match up to another branch moving back towards the North pole, where there will be a turning point given by new, parametrically larger, $\Delta$ and $z_0$. The matching process will allow us to produce a recurrence relation for the sequences $\Delta_n$, $z_n$ of turning points. We can obtain an entirely analytic expression for these values, which, sufficiently close to the horizon, match precisely with the values obtained from numerically integrated surfaces.

\subsection{The linear regime}

We start with the linear approximation, valid for $z,\dot{z}\ll1$. In this case, the minimal surface equation  obtained from \eqref{lagmin} reduces to
\begin{equation}
 \ddot{z}+(d-2)\cot\ph \, \dot{z} -(d-1)\, \xi \, z=0\,, \qquad \xi \equiv \rh \, \kappa. 
\end{equation}
This equation has regular singular points at $\ph=0,\pi$, where the solutions have asymptotics $z\sim \ph^\sigma$ for $\sigma=0,3-d$ (or constant and $z\sim \log\ph$ asymptotics in $d=3$). For $\xi>0$ it is easy to show that a solution regular at $\ph=0$ cannot be regular at $\ph=\pi$, so the general solution can be written
\begin{equation}
 z(\ph)=A \; z_r(\ph)+B\;  z_r(\pi-\ph)
\end{equation}
using the symmetry under $\ph \mapsto\pi-\ph$. Here $z_r$ is chosen by demanding $z_r(\ph)\to1$ as $\ph\to0$, and the asymptotics are characterized by a parameter $\lambda$, by $z_r(\ph)\sim \lambda\, (\pi-\ph)^{3-d}$ as $\ph\to\pi$. In fact, the solutions can be found explicitly in terms of associated Legendre functions, and $\lambda(d,\xi)$ can be found analytically.

\subsection{Near-pole flat space regime}

For the second approximation scheme, we assume $z\ll1,z\ll\dot{z},\ph\ll1$, in which case the problem reduces to that of a spherically symmetric ``soap bubble'' in flat $d$-dimensional Euclidean space, where $(\ph,z)$ are the radial and longitudinal coordinates respectively in a cylindrical polar system. This has the solution
\begin{equation}\label{polesol}
 z=z_0\pm \Delta\int_1^{\frac{\ph}{\Delta}}\frac{\mathrm{d}x}{\sqrt{x^{2d-4}-1}}.
\end{equation}
The constants  $z_0$ and $\Delta$ will be small values, describing the distance from horizon and pole respectively of the turning point of the surface. Note that taking the limit $\ph\to\infty$, the integral converges, so $z$ tends to a fixed value when $d>3$: the upper and lower branches of the surface remain at bounded separation, characterized by the value of
\begin{equation}
 I_d=\int_1^{\infty}\frac{\mathrm{d}x}{\sqrt{x^{2d-4}-1}}=-\sqrt{\pi}\;\frac{\Gamma\left(\frac{d-3}{2(d-2)}\right)}{\Gamma\left(\frac{-1}{2(d-2)}\right)}.
\end{equation}
The asymptotics are given by
\begin{equation}
 z\sim z_0\pm I_d\, \Delta\; \mp\, \frac{1}{d-3}\, \Delta^{d-2}\, \ph^{3-d}\quad \text{ as } \;\;\frac{\ph}{\Delta}\to\infty \,.
\end{equation}
%

\subsection{Matching asymptotics}

With solutions in the two approximation schemes in hand, it now remains only to fix constants of integration by matching the asymptotics. For definiteness, consider starting from a known $(\Delta_n,z_n)$ close to the North pole, computing the coefficients of the linear solution, and finally extracting the values of $(\Delta_{n+1},z_{n+1})$ close to the South pole.

The case $d=3$ is a little different from $d>3$, because of the logarithmic asymptotics, so we will stick to $d>3$ here. A very similar analysis holds in this case, for which we will omit the details, and present only the results.

Recall the form of the linear solution $z(\ph)=A\, z_r(\ph)+B\, z_r(\pi-\ph)$, with the terms regular at $\ph=0,\pi$ respectively. Consistency imposes that the coefficients have parametrically different sizes, with $A\gg B$.  Towards the South pole, the $B$-mode is also suppressed relative to the $A$-mode by the growth of the latter, which renders $B$ essentially irrelevant, and the turning points can be characterized by a single parameter. Indeed, by matching the asymptotics near the South pole, for consistency we need $z_{n+1}=I_d\, \Delta_{n+1}$. Finally, we match $A$ with the relevant coefficients at each end,
\begin{equation}
 z_n+I_d\, \Delta_n=A=\frac{\Delta_{n+1}^{d-2}}{\lambda\, (d-3)},
\end{equation}
which finally gives us our recurrence relation
\begin{equation}
 \Delta_{n+1}=[2\,\lambda \,(d-3)\,\Delta_n]^\frac{1}{d-2},
\end{equation}
demonstrating a parametric growth in $\Delta,z$ at each step.
This can be conveniently solved by changing to the coordinate $x$, defined by $r=r_+ + e^{2\,x}$, (or $\frac{1}{2}\, \rh\, \xi \, z^2 = e^{2\,x}$) from which we get
\begin{equation}
 x_n=\frac{x_0-\nu}{(d-2)^n}+\nu, \qquad \Delta_n=\sqrt{\frac{2}{\kappa}}\; \frac{e^{x_n}}{r_+\, I_d}
\end{equation}
with the constant $\nu$ given by
\begin{equation}
 \nu=\frac{1}{2}\, \log\left(\frac{\rh\, \xi}{2}\right) +\frac{\log[2\,(d-2)\, \lambda \, I_d^{d-2}]}{d-3}.
\end{equation}
It should be noted that the initial value of $x$ where the surface smoothly crosses the pole $\ph=0$, closest to the horizon, is not at $x_0$, but rather at $x=x_0+\log2$, since the solution here is not given by \eqref{polesol}, but by that obtained in the linear regime: constant $x$ to leading order.

For completeness, we record the computed values of $\lambda$ here:
\begin{align} 	
\lambda &= 
\frac{\cosh(\pi\,\j)}{(d-3)\,\pi} \; \;  \frac{2^{d-2}\left(\left(\frac{d-3}{2}\right)!\right)^2}{[(\frac{1}{2})^2+\j^2][(\frac{3}{2})^2+\j^2]\cdots[(\frac{d-4}{2})^2+\j^2]} 
\,, \hspace{4.8cm} d \; \text{odd}
\nonumber \\
\lambda &= (d-3)\,\frac{\sinh(\pi\,\j)}{\j} \left(\frac{1}{1^2+\j^2}\right)\left(\frac{3}{2^2+\j^2}\right)\left(\frac{5}{3^2+\j^2}\right)\cdots\left(\frac{d-5}{\left(\frac{d-4}{2}\right)^2+\j^2}\right)
\,, \qquad d \; \text{even}
\end{align}
where in both cases, $\j$ is defined by
\begin{equation}
\j^2=(d-1)\, \xi-\frac{(d-2)^2}{4}.
\end{equation}

In $d=3$, we have slightly altered formulae:
\begin{align}
 \Delta_{n+1}   &=   -2\,\lambda\,\Delta_n\,\log\left(\frac{\Delta_n}{2}\right) \\ 
 x_n   &=   \frac{1}{2}\log\left(\frac{\rh\,\xi}{2}\right) +\log\left[-\Delta_n\log\left(\frac{\Delta_n}{2}\right)\right] \\
 \lambda  &=  \frac{2\cosh(\pi\,\j)}{\pi}\,.
\end{align}
%

\subsection{Validity}
Having done this calculation, it is worth checking to see that we have done something sensible: there must be some overlap between the domains of validity of each approximation. It is straightforward to confirm this \emph{a posteriori}, looking at overlaps near the North pole for definiteness. Firstly, the linear approximation is valid when $z,\dot{z}\ll1$, which requires $\ph^{2-d}$ to be much smaller than the coefficient of the growing mode. This holds for $\ph\gg \Delta$. Secondly, the near-pole approximation is valid when $z\ll\dot{z}$, which requires $\ph^{d-2}\ll\Delta^{d-3}$ (as well as $\ph\ll1$). This means that as long as $\Delta$ remain small, both approximations are simultaneously valid for
\begin{equation}
 \Delta\ll\ph\ll\Delta^\frac{d-3}{d-2}.
\end{equation}
For $d=3$, we have a similar result, with overlap region
\begin{equation}
 e^{-\frac{1}{\Delta}}\ll\ph\ll \frac{-1}{\log\Delta}\,.
\end{equation}
%

\subsection{Relation to $\phA$}

We can straightforwardly use this analysis to show that for $\phA$ sufficiently close to $\pi/2$ there will be an infinite tower of minimal surfaces. This is because each branch of a surface is characterized by a single parameter. All surfaces with a particular turning point will match closely outside that turning point. To obtain a surface on later branches of this tower requires the initial value of $-x$ to be exponentially large, so the distance from the horizon will be suppressed by an exponential of an exponential.

Incidentally, this gives an indirect demonstration of non-existence of connected minimal surfaces for sufficiently large $\phA$. The connected surfaces form a one-parameter family, parameterized by the maximum depth reached into the bulk, or equivalently the location $x_0$ of the surface at $\ph=0$. Considering $\phA$ as a function of $x_0$, continuity implies that there are only two ways for $\phA$ to reach arbitrarily close to $\pi$. The first is for $\ph_\infty=\pi$ for some $x_0$, but this is precluded from happening by behaviour near the boundary. The other way is for a sequence $x_m$ to exist such that $\phA(x_m)$ tends to $\pi$, but the self-similarity properties of surfaces indicated here will prevent this option also.

%

\providecommand{\href}[2]{#2}\begingroup\raggedright\endgroup

\end{document}